\documentclass{elsart}
\usepackage{epsfig}
\usepackage{mathrsfs}                                                                                
\begin{document}
                                                                                
\begin{frontmatter}

\title{A relativistic two-nucleon model for $A(p,K^+){_\Lambda}B$
reaction}

\author[Saha]{R. Shyam}
\author[Giessen]{H. Lenske}
\author[Giessen]{U. Mosel}

\address[Saha]{Saha Institute of Nuclear Physics, Kolkata,  India.}
\address[Giessen]{Institute f\"ur Theoretische Physik, Universit\"at Giessen,
Giessen, Germany}

\begin{abstract}
We investigate the $A(p,K^+){_\Lambda}B$ reaction within a covariant two 
nucleon model. We focus on amplitudes which are described by creation, 
propagation and decay into relevant channel of $N^*$(1650), $N^*$(1710),
and $N^*$(1720) intermediate baryonic resonance states in the initial
collision of the projectile nucleon with one of its target counterparts
This collision is modeled by the exchange of $\pi$ as well as $\rho$ and
$\omega$ mesons. The bound state nucleon and hyperon wave functions are
obtained by solving the Dirac equation with appropriate scalar and vector
potentials. Expressions for the reaction amplitudes are derived taking
continuum particle wave functions in both distorted wave and plane wave 
approximations. Detailed numerical results are presented in the plane wave 
approximation; estimates of the effects of the initial and final state
interactions are given in the eikonal approximation. Cross sections of
1 - 2 nb/sr at the peak positions, are obtained for favored transitions in
case of heavier targets. 
\end{abstract}


\end{frontmatter}

\newcommand{\tauiso}{\mbox{\boldmath$\tau$}}
\newcommand{\gmu}{\gamma_\mu}
\newcommand{\gf}{{\gamma_5}}                                                                                
\section{Introduction}
$\Lambda$ hypernuclei are the most familiar hypernuclear systems 
which have been studied extensively by stopped as well as in-flight
$(K^-,\pi^-)$ reaction~\cite{chr89,ban90,may81} and also by $(\pi^+,K^+)$
reaction \cite{pil91,hot01}. The kinematical properties of the $(K^-,\pi^-)$
reaction allow only a small momentum transfer to the nucleus (at forward
angles), thus there
is a large probability of populating $\Lambda$-substitutional states
($\Lambda$ assume the same orbital angular momentum as that of the 
neutron being replaced by it). On the other
hand, in the $(\pi^+,K^+)$ reaction the momentum transfer is larger than
the nuclear Fermi momentum. Therefore, this reaction can populate states
with the configuration of an outer neutron hole and a $\Lambda$ hyperon
in a series of orbits covering all the bound states. During past years
data on the hypernuclear spectroscopy have been used extensively to
extract information about the hyperon-nucleon interaction (which could be
quite different from the nucleon-nucleon interaction) within a variety of
theoretical approaches (see, e.g.,~\cite{hiy00,kei00}).

Alternatively, $\Lambda$-hypernuclei can also be produced with high intensity 
proton beams via $p + A(N,Z) \rightarrow {_{\Lambda}}B(N-1,Z) + n + K^+$,
$p + A(N,Z) \rightarrow {_{\Lambda}}B(N,Z-1) + p^\prime + K^+$, and
$p + A(N,Z) \rightarrow {_{\Lambda}}B(N,Z) + K^+$ reactions where $N$ and
$Z$ are the neutron and proton numbers, respectively, in the target
nucleus. In this paper we study the last reaction [to be referred to as
$A(p,K^+){_{\Lambda}}B$] where the hypernucleus ${_{\Lambda}}B$ has the
same neutron and proton numbers as the target nucleus $A$, with one
hyperon added. This reaction
is exclusive in the sense that the final channel is a two body system.
First set of very preliminary data have already been reported
for this reaction on deuterium and helium targets~\cite{kin98}. More
measurements for this reaction involving also the heavier targets are
expected to be performed at the COSY facility of the Forschungszentrum
J\"ulich [see, e.g., Ref.~\cite{sch95}]. This reaction involves much
larger momentum transfer to the nucleus as compared to the
$(\pi^+,K^+)$ reaction [more than 1.0 GeV/c (see Fig.~1) as compared to
only 0.33 GeV/c at the outgoing $K^+$ angle of 0$^\circ$]. Therefore,
it samples bound state wave functions in a region where they are very
small and are unlikely to be reached in other reactions. From the 
spectroscopic point of view, the states of the hypernucleus $_{\Lambda}B$
excited in the $(p,K^+)$ reaction may have a different type of 
configuration as compared to those reached in the $(\pi^+,K^+)$
reaction.  Thus a comparison of informations extracted from 
the study of two reactions is likely to provide a better
understanding of the hypernuclear structure.
\begin{figure}
\begin{center}
\mbox{\epsfig{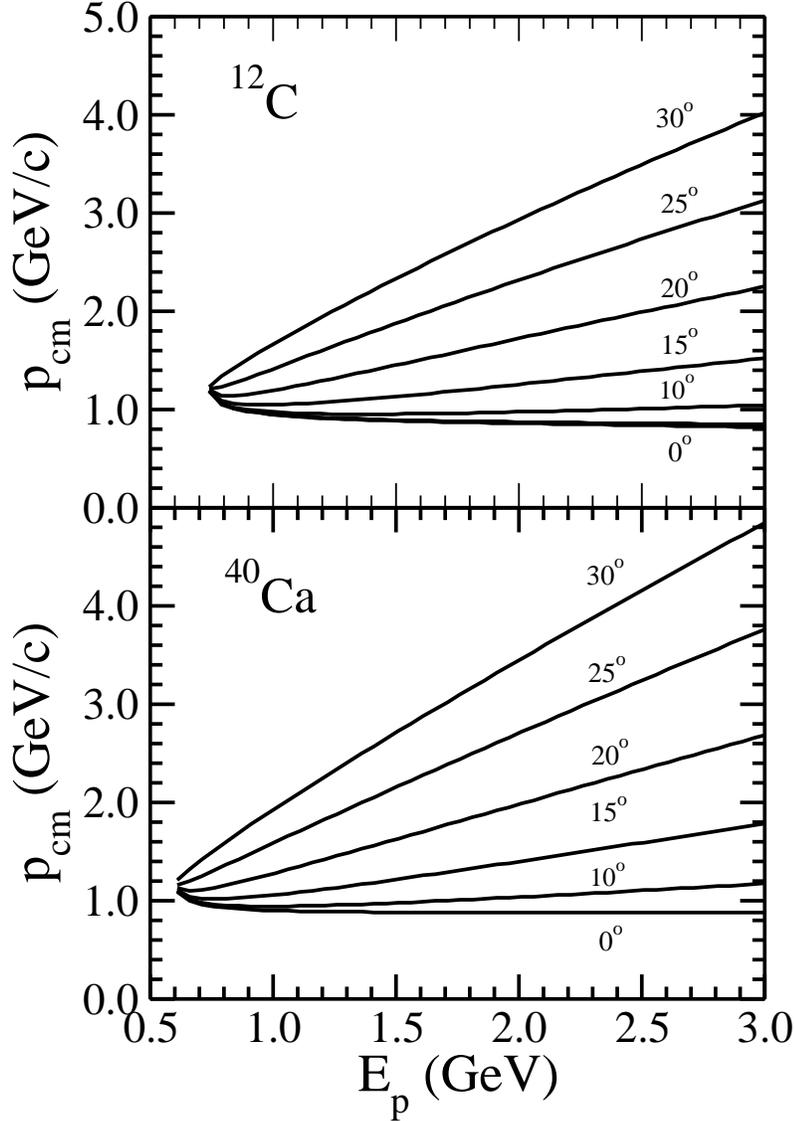}}
\end{center}
\caption{Center-of-mass momentum transfer available to the ($p,K^+$)
reaction as a function of the laboratory bombarding energy ($E_p$) for
$^{12}$C and $^{40}$Ca targets. For a fixed energy the allowed kinematic
region extends from 0$^\circ$ to 180$^\circ$. We have shown here only a few
angles.
}
\label{fig:figa)}
\end{figure}

Theoretical studies of the $A(p,K^+){_{\Lambda}}B$ reaction are rather
sparse and preliminary in nature~\cite{shi86,kom95,fet98,kri95}.
They are based on two main approaches; the one-nucleon model (ONM)
[Fig.~2(a)] and the two-nucleon model (TNM) [Figs.~2(b) and 2(c)]. In
the ONM the incident proton first scatters from the target nucleus
and emits a (off-shell) kaon and a $\Lambda$ hyperon. Subsequently,
the kaon rescatters into its mass shell while the hyperon  gets 
captured into one of the (target) nuclear orbits. Thus there is only
a single active nucleon (impulse approximation) which carries the entire
momentum transfer to the target nucleus. This makes this model extremely
sensitive to details of the bound state wave function at very large
momenta where its magnitude is very small leading to quite low cross
sections. In the ONM calculations of $(p,K^+)$ and also of $(p,\pi)$
reactions the distortion effects in the incident and the outgoing channels
have been found to be quite important~\cite{kri95,fea79,coo82}. 
\begin{figure}
\begin{center}
\mbox{\epsfig{file=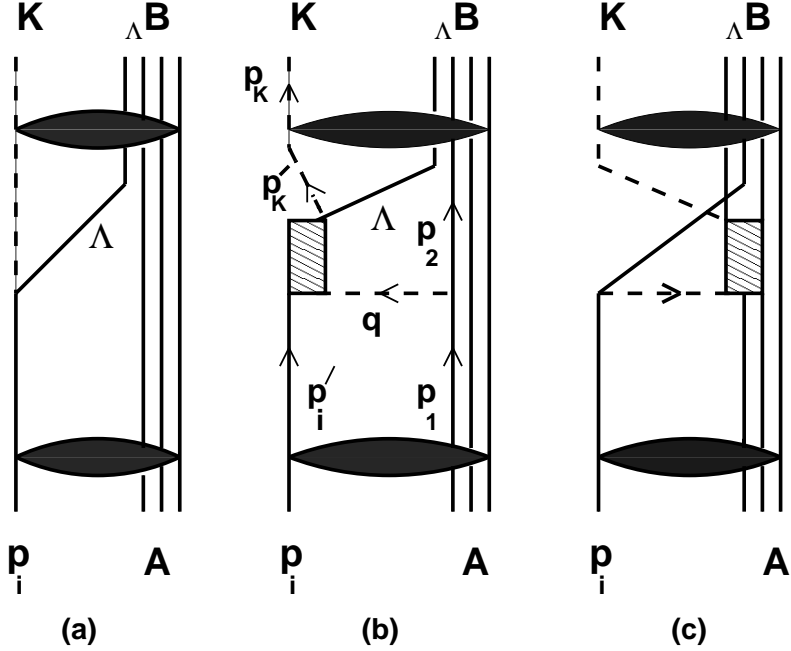,width=.75\textwidth}}
\end{center}
\caption{Graphical representation of one-nucleon (a) and two-nucleon 
[(b) and (c)] models. The elliptic shaded area represent the optical 
model interactions in the incoming and outgoing channels.  
}
\label{fig:figb)}
\end{figure}

In the two-nucleon mechanism, on the other hand, the kaon
production proceeds via a collision of the projectile nucleon
with one of its target counterparts. This excites intermediate baryonic 
resonance states which decay into a kaon and a $\Lambda$ hyperon. 
The nucleon and the hyperon are captured into the respective nuclear 
orbits while the kaon rescatters into its mass shell. In this picture
there are altogether three active bound state baryon wave functions
taking part in the reaction process, thus allowing the large momentum transfer
to be shared among three baryons. Consequently, the sensitivity of the
model is shifted from high momentum parts of the bound state wave
functions (not very well known) to those at relatively lower momenta
where they are rather well known from $(e, e^\prime p)$ and $(\gamma, p)$
experiments and are relatively larger [see, e.g.,~\cite{fru84}]. This
could lead to larger cross sections. Moreover, in the TNM studies of other
large momentum transfer reactions like $(p,\pi)$, $(\gamma,\pi)$, and
$(\gamma,\eta)$ the distortion effects have been found to be less
pronounced~\cite{alo88,pet98}.

Most of the previous calculations~\cite{shi86,kom95,kri95} of the
$A(p,K^+){_{\Lambda}}B$ reaction have been done within the non-relativistic
framework.  However, for processes involving momentum transfers of
typically 400 MeV/c or more, a non-relativistic treatment of corresponding
wave functions is questionable as in this region the lower component of the
Dirac spinor is no longer negligible in comparison to its upper component
(see, e.g, Ref.~\cite{shy95}). Furthermore, in the non-relativistic
description one must resolve the ambiguities of the pion-nucleon-nucleon
vertex that result from the non-relativistic reduction of the full covariant
pion-nucleon-nucleon vertex (see, e.g., \cite{alo88,shy95,bar69} for more
details). This is important as this vertex is used to describe the initial
collision between the incoming proton and the target nucleus.

In this paper, we study the $A(p,K^+){_{\Lambda}}B$ reaction within a
fully covariant TNM by retaining the field theoretical structure of the 
interaction vertices and by treating the baryons as Dirac particles.
The initial interaction between the incoming proton and a bound nucleon of 
the target is described by the $\pi$, $\rho$ and $\omega$  exchange 
diagrams which were also used to describe the $K^+$ production in
elementary proton-proton ($pp$) collisions~\cite{shy99,shy01,shy04a}.
In a previous study of the $(p,K^+)$ reaction on a $^{40}$Ca
target~\cite{shy04} the latter two meson exchanges were not included. 
Some authors, however, have used an alternative approach for the elementary
$pp \to pK^+\Lambda$ reaction~\cite{kai99,gas00,lag91} in which kaon
exchange mechanism between the two initial state nucleons leads to
the $K^+$ production. We have not considered this mechanism here. There 
are indications, from from some results reported by COSY-TOF and ANKE
groups~\cite{sch03,bus04}, that the Dalitz plots for the
$pp \to pK^+\Lambda$ reaction are dominated by the excitation of
nucleon isobars, though modified by the $\Lambda p$ final state
interaction (FSI) (see also Ref.~\cite{wil04}).

In our model, the initial state interaction of the incoming proton 
with a bound nucleon leads to $N^*$(1650)[$\frac{1}{2}^-$],
$N^*$(1710)[$\frac{1}{2}^+$], and $N^*$(1720)[$\frac{3}{2}^+$] baryonic
resonance intermediate states  which make predominant contributions to
elementary $pp \to p K^+ \Lambda$ cross sections in the beam energy regime 
of near threshold to 10 GeV \cite{shy99}. Terms corresponding to
the interference among various resonance excitations are included in the
total reaction amplitude. We have ignored the diagrams of type 2(a) since
contributions of such processes are expected to be very small in comparison
to those of the TNM.

In section II, we present the details of our formalism for calculating 
amplitudes corresponding to the diagrams shown in Figs. 2b and 2c, using
continuum wave functions within both distorted wave (DW) and plane wave 
(PW) approximations. In section III, numerical results are presented
for $^{4}$He$(p,K^+)$$^{5}\!\!\!_\Lambda$He,
$^{12}$C$(p,K^+)$$^{13}\!\!\!_\Lambda$C, and
$^{40}$Ca$(p,K^+)$$^{41}\!\!\!_\Lambda$Ca reactions within the 
the PW approximation. We also give here the estimates of the
effects of the initial and final state interactions which are
calculated within the eikonal approximation. Summary, conclusions 
and future outlook of our work are
given in section IV. Finally some formulas for the amplitudes 
are summarized in appendix A.

\section{Covariant two-nucleon model for $A(p,K^+)$ reactions}
\subsection{Kinematics, elementary processes and contributing diagrams}

The elementary associate production reaction $pp \to pK^+\Lambda$ has been 
studied extensively both experimentally and theoretically (see, e.g.,
Ref.~\cite{mos02} for a recent review). The laboratory threshold energy for
this reaction is 1.582 GeV. For the case of the $A(p,K^+){_\Lambda B}$
reaction the threshold energy is given by 
\begin{eqnarray}
E_p^{thres} & = & \frac{(m_B^2-m_A^2)}{2m_A} + \frac{(m_K^2-m_N^2)}{2m_A}
                  + \frac{m_Bm_K}{m_A} - m_N,
\end{eqnarray}
where $m_B$ and $m_A$ are the masses of the hypernucleus $_\Lambda B$
and the target nucleus A, respectively and $m_K$ is the mass of the
kaon. $E_p^{thres}$ is 0.739 GeV and 0.602 GeV for  $^{12}$C and $^{40}$Ca
targets, respectively which are considerably lower than the threshold
energy value of the elementary production process as stated above.

The $(p,K^+)$ reaction is characterized by the momentum transfer
${\bf p}_{cm} = {\bf p}_i - {\bf p}_K$ to the residual nucleus. In Fig.~1,
${p}_{cm}$ are plotted as a function of the laboratory beam
energy for several values of the outgoing kaon angles for $^{12}$C and
$^{40}$Ca targets. It is clear that $p_{cm}$ varies between 1 GeV to 5 GeV.
In the ONM a single nucleon alone carries this huge momentum to
the nucleus. In the PW approximation, the ONM cross section of the
$(p,K^+)$ reaction is directly proportional to the modulus square of the
bound state wave functions in the momentum space. Since at these large
momenta the magnitudes of the bound state wave functions are very small,
the corresponding cross sections are quite weak.  Of course, the inclusion
of the distortions changes this model
towards a multi-nucleon mechanism thus shifting the sensitivity of the
model to lower momentum components of the bound state wave functions.
Therefore, {\it apriori}, these effects are very important in the ONM
description of such reactions.

The data on the associate production reaction $pp \to p\Lambda K^+$, are
available for beam energies ranging from very close to
the threshold to upto 10 GeV which can be well described within an
effective Lagrangian approach~\cite{shy99,shy01,shy04a}. In this model,
the associated kaon production proceeds via excitation, propagation and
decay of the $N^*$(1650), $N^*$(1710) and $N^*$(1720) intermediate baryonic
resonant states, in the initial collision of two nucleons in the 
incident channel. The coupling constants at the nucleon-nucleon-meson
vertices are fixed by fitting to an
elastic $NN$ $T$ matrix while those at resonance-nucleon-meson vertices
are determined from the experimental branching ratios for the decay of
the resonance into relevant channels except for those involving the $\omega$
meson where they are determined from the vector meson dominance (VMD)
hypothesis. To describe the near threshold data
\begin{figure}
\begin{center}
\mbox{\epsfig{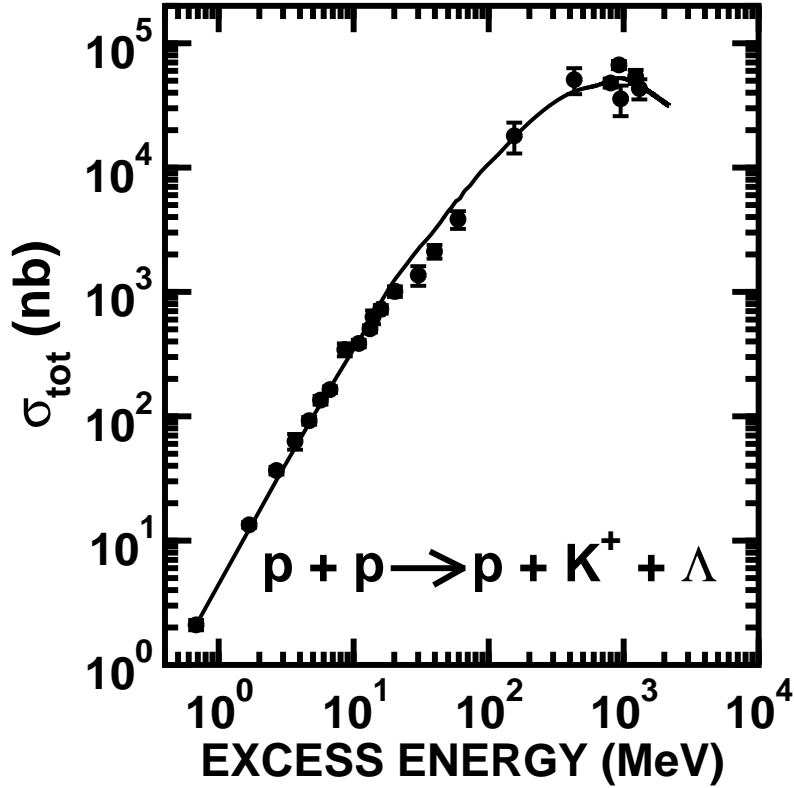}}
\end{center}
\caption{Comparison of the calculated total cross sections for the 
$p + p \rightarrow p + K^+ + \Lambda$ reaction with the corresponding 
experimental data [taken from Refs.~\protect\cite{lan88,kow04}] as 
a function of the excess energy. The calculated cross sections represent
the coherent sum of the amplitudes corresponding to all the contributing
baryonic resonant states and the meson exchange processes. 
}
\label{fig:figc)}
\end{figure}
\noindent
the FSI effects in the final channel are included within the framework
of the Watson-Migdal theory. As an example, we show in Fig.~3 a comparison
of calculated and experimental total cross sections for the
$pp \to pK^+\Lambda$ reaction. The FSI amplitude has been calculated
using the scattering length and the effective range parameters of the 
model $\tilde{A}$ of $\Lambda-p$ interaction described in Ref.~\cite{reu94}.
The cross sections are plotted as a function of the excess energy (defined
as $\sqrt{s} - m_N - m_{K^+} - m_{\Lambda}$, where $\sqrt{s}$ is the
invariant mass). It is clear that the model provides a good description of
the experimental data in the entire range of beam energies. While at near
threshold beam energies the reaction proceeds predominantly via excitation
of the $N^*$(1650) resonance, the other resonances become more and more 
important with increasing beam energies. One pion exchange graphs
dominate the production process for all the energies.

The structure of our TNM for the $(p,K^+)$ reaction is similar to that
of the effective Lagrangian approach described above. The 
initial interaction between the incoming proton and a bound
nucleon of the target is described not only by the dominant one-pion
exchange mechanism but also by the $\rho$ and $\omega$ exchange
processes. The latter may become more important at backward angles
of the outgoing $K^+$ spectrum where momentum transfers are relatively
larger. We use the same effective Lagrangians and vertex parameters to
model these interactions. The initial state interaction between
the two nucleons leads to the $N^*(1650)[\frac{1}{2}^-]$,
$N^*(1710)[\frac{1}{2}^+]$, and $N^*(1720)[\frac{3}{2}^+]$
baryonic resonance intermediate states. The vertex parameters 
here too are the same as those used in the description of the elementary
reaction. Terms corresponding to the interference between various
amplitudes are retained.

We denote the diagrams (and the corresponding amplitudes) by "projectile
emission (PE)" if the intermediate meson starts from the projectile and
rescatters from the target nucleon which is excited to an
intermediate baryonic resonance state [Fig.~2(c)] and by "target emission"
if the opposite is the case [Fig.~2(b)]. In the two amplitudes 
there is a tremendous variation in the energy of the intermediate meson.
For example, in Fig.~2(b), the energy of the exchanged meson is given by
the difference between single-particle binding energies of the initial
and final intermediate states which is very small and can be approximated
to zero. On the other hand, in Fig.~2(c), the energy of the intermediate 
meson is almost equal to the incident proton energy. Due to different
four momentum transfers the behavior of the two amplitudes is quite 
different and they explore different regions of the intermediate meson
propagator. Thus corresponding medium effects have to be known over a 
wide range of energy and momenta in order to incorporate their effects
into the calculations. 
\subsection{Calculations of the TNM amplitudes}
 
The effective Lagrangian for the nucleon-nucleon-meson 
vertices are given by
\begin{eqnarray}
{\mathscr L}_{NN\pi} & = & -\frac{g_{NN\pi}}{2m_N} {\bar{\Psi}} \gamma _5
                             {\gamma}_{\mu} \tauiso
                            \cdot (\partial ^\mu {\bf \Phi}_\pi) \Psi. \\
{\mathscr L}_{NN\rho} & = &- g_{NN\rho} \bar{\Psi}_N \left( \gmu + \frac{k_\rho}
                         {2 m_N} \sigma_{\mu\nu} \partial^\nu\right)
                          \tauiso \cdot \mbox{\boldmath $\rho$}^\mu \Psi_N. \\
{\mathscr L}_{NN\omega} & = &- g_{NN\omega} \bar{\Psi}_N \left( \gmu +
                         \frac{k_\omega}
                         {2 m_N} \sigma_{\mu\nu} \partial^\nu\right)
                          \omega^\mu \Psi_N. 
\end{eqnarray}
We have used notations and conventions of Bjorken and Drell~\cite{bjo64}.
In Eq.~(2) $m_N$ denotes the nucleon mass. In Eqs.~(3) - (4)
$\sigma_{\mu\nu}$ is defined as
\begin{eqnarray}
\sigma_{\mu\nu} & = & { i \over 2}(\gamma_\mu\gamma_\nu - \gamma_\nu\gamma_\mu)
\end{eqnarray}
The $NN$-meson vertices are corrected for off-shell effects by
multiplying the corresponding coupling constants by form factors of the
forms~\cite{shy99}
\begin{eqnarray}
F_i^{NN} & = & \left( \frac{\lambda_i^2 - m_i^2}{\lambda_i^2 - q_i^2} \right),
\end{eqnarray} 
where $q_i$ and $m_i$ are the four momentum and mass of the $i$th exchanged
meson and $\lambda_i$ are the corresponding cutoff parameters [their values
are taken to be the same as those given in Ref.~\cite{shy99}]. The latter
governs the range of the suppression of the contributions of 
high momenta.

As the $\Lambda$-hyperon has zero isospin, only isospin-1/2 nucleon
resonances are allowed. Below 2 GeV center of mass (c.m.) energy, only
three resonances, $N^*$(1650), $N^*$(1710), and $N^*$(1720),
have significant decay branching ratios into the $K^{+}\Lambda$ channel.
As in the description of the associate $K^+$ production in the
elementary $pp$ collisions, only these three resonances have been
considered in this work. The effective Lagrangians for the
resonance-nucleon-meson vertices are written
as~\cite{shy99,benm95,benm89,feus97,feus98}
\begin{eqnarray}
{\mathscr L}_{N_{1/2}^*N\pi} & = & -g_{N_{1/2}^*N\pi}
                          {\bar{\Psi}}_{N_{1/2}^*} i{\Gamma} \tauiso
                        {\bf \Phi}_\pi \Psi 
                          + {\rm h.c.},\\
{\mathscr L}_{N_{1/2}^*N\rho} & = &- g_{N_{1/2}^*N\rho} \bar{\Psi}_{N^*}
                           \frac{1} {2m_N}
                            \Gamma_{\mu\nu} \partial^\nu
                      \tauiso \cdot \mbox{\boldmath $\rho$}^\mu \Psi_N. +
                         {\rm h.c.}\\
{\mathscr L}_{N_{1/2}^*N\omega} &=&- g_{N_{1/2}^*N\omega} \bar{\Psi}_{N^*}
                         \frac{1}{2m_N}
                         \Gamma_{\mu\nu} \partial^\nu
                          \omega^\mu \Psi_N. + {\rm h.c.}  \\
{\mathscr L}_{N_{3/2}^*N\pi} & = & \frac{g_{N_{3/2}^*N\pi}}{m_\pi}
                         {\bar{\Psi}}_{\mu}^{N^*} \Gamma_\pi
                         {\tauiso} \cdot \partial ^{\mu}
                         {\bf \Phi}_\pi \Psi + {\rm h.c.}.\\
{\mathscr L}_{N_{3/2}^*N\rho} & = & {\rm i} \frac{g_{N_{3/2}^*N\rho}}
                        {m_{N^*} + m_N}{\bar{\Psi}}_{\mu}\tauiso \left(
                        \partial^\nu \mbox{\boldmath $\rho$}^\mu -
                        \partial^\mu \mbox{\boldmath $\rho$}^\nu
                        \right) \gamma_\nu \gf\Psi _N + {\rm h.c.}.\\
{\mathscr L}_{N_{3/2}^*N\omega} & = & {\rm i} \frac{g_{N_{3/2}^*N\omega}}
                        {m_{N^*} + m_N}{\bar{\Psi}}_{\mu} \tauiso \left(
                        \partial^\nu \omega^\mu -
                        \partial^\mu \omega^\nu \right)
                         \gamma_\nu \gf\Psi _N + {\rm h.c.}.
\end{eqnarray}
The operator $\Gamma(\Gamma_\pi)$ is either $\gamma_5$ (unity) or unity
$(\gamma_5)$ depending on whether the parity of the resonance is even
or odd, respectively. The operator $\Gamma_{\mu\nu}$ is
$\sigma_{\mu\nu}$ or $\gamma_5 \sigma_{\mu\nu}$ for these two cases.
${\bar {\Psi}}_\mu$ is the $N^*$(1720) vector spinor. It may be noted
that we have used the conventional Rarita-Swinger coupling for the
spin-$\frac{3}{2}$ particle~\cite{benm95,benm89} which also has
off-shell contributions to the spin-$\frac{1}{2}$ partial waves. To
describe the latter, an operator
$\Theta_{\alpha\mu}(z) = g_{\alpha\nu} -\frac{1}{2}(1+2z)\gamma_\alpha
\gamma_\mu$ has also been included in the vector spinor vertex 
\cite{benm95,feus97,feus98,pen02}. The choice of the off-shell parameter
$z$ is arbitrary; it is treated as a free parameter to be determined
by fitting to the data. In Ref.~\cite{pas98} an alternative interaction
has been suggested that does not activate the spurious spin-$\frac{1}{2}$
degree of freedom.  We, however, work with the Lagrangians as given in
Eqs.~(10)-(12). It may further be noted that we have used a 
pseudovector (PV) coupling for the $NN\pi$ vertex and a pseudoscalar
(PS) one for $N_{1/2}^*N\pi$ and $N_{1/2}^*\Lambda K$ vertices. This
provides the best description of the $pp \to p\Lambda K^+$ data~\cite{shy99}.
 
The effective Lagrangians for the resonance-hyperon-kaon vertices are
written as 
\begin{eqnarray}
{\mathscr L}_{N_{1/2}^*\Lambda K^+} & = & -g_{N^*_{1/2}\Lambda K^+}
                          {\bar{\Psi}}_{N^*} {i\Gamma} \tauiso
                        {\bf \Phi}_{K^{+}} \Psi
                        + {\rm h.c.}.\\
{\mathscr L}_{N_{3/2}^*\Lambda K^+} & = & \frac{g_{N^*_{3/2}\Lambda K^+}}
                         {m_{K^+}}
                         {\bar{\Psi}}_{\mu}^{N^*} \Gamma_\pi
                         {\tauiso} \cdot \partial ^{\mu}
                         {\bf \Phi}_{K^{+}} \Psi + {\rm h.c.}.
\end{eqnarray}
Signs and values of various coupling constants have been taken from
\cite{shy99,shy04} and are shown in Table 1. It should be noted that
for the nucleon-nucleon-meson vertices, the coupling constants were
taken to be energy dependent in the same way as that described in Ref.
\cite{shy99}. The $N_{1/2}^*N$-meson and $N_{3/2}^*N$-meson vertices are
corrected for the off-shell effects by multiplying the corresponding
coupling constants by form factors of the forms (which is also used in
Ref.~\cite{shy03}).
\begin{eqnarray}
F_j^{NN^*} & = & \left[ \frac{(\lambda_i^{N^*})^4}{(\lambda_j^{N^*})^4 +
               (q_j^2 - m_j^2)^2} \right], j = \pi, \rho, \omega
\end{eqnarray}
The values of the cut-off parameters appearing therein are taken to be 1.2 GeV
in all the cases in order to reduce the number of free parameters.
It may noted that same form of $F_j^{NN^*}$ and the value of the cut-off
parameter was also used in the calculations of the elementary cross 
sections shown in Fig.~3.
 
After having established the effective Lagrangians and the coupling
constants, one can  write down, by following the well known
\begin{table}
\begin{center}
\caption [T1]{Coupling constants for various vertices
used in the calculations.
}
\vspace{1.0cm}
\begin{tabular}{|c|c|}\hline
Vertex & Coupling Constant($g$)\\ 
       &                \\
\hline
$NN\pi$                & 12.56 \\
$NN\rho$               & 2.00  \\
$NN\omega$             & 24.05 \\
$N^*(1710)N\pi$        & 1.04  \\
$N^*(1710)N\rho$       & 4.14  \\
$N^*(1710)N\omega$     & 1.22  \\ 
$N^*(1710)\Lambda K^+$ & 6.12  \\
$N^*(1650)N\pi$        & 0.81  \\
$N^*(1650)N\rho$       & 2.62  \\
$N^*(1650)N\omega$     & 1.80  \\ 
$N^*(1650)\Lambda K^+$ & 0.76  \\
$N^*(1720)N\pi$        & 0.21  \\
$N^*(1720)N\rho$       & 33.75 \\
$N^*(1720)N\omega$     & 16.93 \\
$N^*(1720)\Lambda K^+$ & 0.87  \\
\hline
\end{tabular}
\end{center}
\end{table}
Feynman rules, the amplitudes for graphs 2(b) and 2(c). The isospin
part is treated separately which gives rise to a constant factor for
each graph. For example, the amplitude for graph 2(b) with one-pion exchange
mechanism and excitation of the positive parity spin-${1 \over 2}$ baryonic
resonance is given by,
\begin{eqnarray}
M_{2b}(N^*_{1/2}) & =& C_{iso}^{2b}\biggl(\frac{g_{NN\pi}}{2m_N}\biggr)
(g_{N_{1/2}^*N\pi}) (g_{N^*_{1/2}\Lambda K^+})
{\bar{\psi}}(p_2) \gamma _5 \gamma_\mu q^\mu \nonumber \\
& \times & \psi(p_1) D_{\pi}(q)
{\bar{\psi}}(p_\Lambda)\gamma_5 D_{N^*_{1/2}}(p_{N^*}) \gamma_5 \nonumber \\
& \times & \Phi^{(-)*}_{K}(p_K^\prime, p_K)\Psi^{(+)}_i(p_i^\prime, p_i),
\end{eqnarray}
where various momenta are as defined in Fig.~2(b). In addition, $p_{N^*}$
is the momentum associated with the intermediate resonance and $p_\Lambda$
is that associated with the $\Lambda$ hyperon. The isospin factor 
$C_{iso}^{2b}$ is unity. For the graph 2(c) also it is unity. The
functions $\psi$ are the four component (spin space) Dirac spinor in
momentum space~\cite{shy95,shy91}. $\Phi_{K}^{(-)*}(p_K^\prime, p_K)$
[$\Psi_i^{(+)}(p_i^\prime, p_i)$] is the wave function for the outgoing
kaon [incoming proton] with appropriate boundary conditions. Function
$D_i(p_i)$ is the propagator for the $i${\it th} particle with momentum
$p_i$. The propagators for mesons and spin-${1 \over 2}$ and
spin-${3 \over 2}$ intermediate resonances are given by  
\begin{eqnarray}
D_\pi(q) & = & {i \over {q^2 - m_\pi^2 -\Pi_\pi(q)}},\\
D_\rho^{\mu\nu}(q) & = & -i\left({g^{\mu\nu}-q^\mu q^\nu/q^2}
                          \over {q^2 - m_\rho^2-\Pi_\rho(q)} \right),\\
D_{N^*_{1/2}} (p) & = & i\left[ {p_\eta\gamma^\eta + m_{N^*_{1/2}}} \over
                       {p^2 - (m_{N^*_{1/2}}-i\Gamma_{N^*_{1/2}}/2)^2}
                         \right ],\\
D_{N^*_{3/2}}^{\mu \nu} (p) & = & -\frac{i(p\!\!\!/ + m_{N^*_{3/2}})}
                    {p^2 - (m_{N^*_{3/2}}-i\Gamma_{N^*_{3/2}}/2)^2}
                               \nonumber \\
                 &   &\times [g^{\mu \nu} - \frac{1}{3}\gamma^\mu \gamma^\nu
                              - \frac{2}{3m_{N^*_{3/2}}^2} p^\mu p^\nu
                              + \frac{1}{3m_{N^*_{3/2}}^2}
                                ( p^\mu \gamma^\nu - p^\nu \gamma^\mu )].
\end{eqnarray}
In Eq.~(17) $\Pi_\pi(q)$ is the (complex) pion self energy which accounts
for the medium effects on the propagation of the pion in the nucleus.
Similarly, $\Pi_\rho(q)$, in Eq.~(18) is the same for the $\rho$ meson.
We have adopted the approach of Ref.~\cite{dmi85} for the calculation
of $\Pi_\pi(q)$ which has been renormalized by including the short-range
repulsion effects through the constant Landau-Migdal
parameter $g^\prime$. All the relevant formulas for this calculation
are given in Ref.~\cite{shy95}.
Like $(p,\pi)$~\cite{shy95} and $(p,p^\prime \pi)$~\cite{jai88}
reactions, $(p,K^+)$ cross sections too are sensitive to the parameter
$g^\prime$~\cite{shy04}. For the self energies of $\rho$ and $\omega$
mesons, we have followed the procedure described in Ref.~\cite{muh04}. 

In Eqs.~(19) and (20), $\Gamma_{N^*}$ is the total width of the resonance
which is introduced in the denominator term to account for the
finite life time of the resonances for decays into various
channels. $\Gamma_{N^*}$ is a function of the center of mass momentum
of the decay channel, and it is taken to be the sum of the widths for pion
and rho decay (the other decay channels are considered only implicitly by
adding their branching ratios to that of the pion channel)~\cite{shy99}.
The medium corrections on the intermediate resonance widths have not been
included. We do not expect any major change in our results due to
these effects. As is pointed out in~\cite{lut03,mal02,eff97}, the medium
correction effects on  widths of the $s$- and $p$-wave resonances, which
make the dominant contribution to the cross sections being investigated
here, are not substantial. On the other hand, any medium modification in
the width of the $d$-wave resonance is unlikely to alter our results as
their contributions to the cross sections are negligible~\cite{muh04}. 

The four component Dirac spinors $\psi(p_j)$ are the solutions of 
the Dirac equation in momentum space for a bound state problem 
in the presence of an external potential field~\cite{shy95,bet57}
\begin{eqnarray}
p\!\!\!/\psi(p) & = & m_N\psi(p) + F(p),
\end{eqnarray}
where
\begin{eqnarray}
F(p) & = & \delta(p_0 - E) \Biggl[\int d^3p^\prime V_s(-{\bf p}^\prime)
\psi({\bf p} + {\bf p}^\prime) \nonumber \\ & - & \gamma_0
 \int d^3p^\prime V_v^0(-{\bf p}^\prime) \psi({\bf p} + {\bf p}^\prime)
                   \Biggr] .
\end{eqnarray}
In our notation '$p$' represents a four momentum, and '${\bf p}$' a three
momentum. The magnitude of ${\bf p}$ ($|p|$) is represented by $k$, and its
directions by ${\hat p}$. '$p_0$' represents the time-like component of
the momentum. Similarly, the magnitude and directions of the radial 
vector ${\bf r}$ is represented by $a$ and ${\hat r}$, respectively.
In Eq.~(22), $V_s$ and $V_v^0$ represent a
scalar potential and time-like component of a vector potential in the
momentum space. The spinors $\psi(p)$ and $F(p)$ are written as
\begin{eqnarray}
\psi(p) & = & \delta(p_0-E){{f(k) {\mathscr Y}_{\ell 1/2 j}^{m_j} (\hat p)}
                    \choose {-ig(k){\mathscr Y}_{\ell^\prime 1/2 j}^{m_j}
                     (\hat p)}}, \nonumber \\ 
F(p) & = & \delta(p_0-E){{\zeta(k) {\mathscr Y}_{\ell 1/2 j}^{m_j} (\hat p)}
                \choose {-i\zeta^\prime(k){\mathscr Y}_{\ell^\prime 1/2 j}
                     ^{m_j} (\hat p)}},  
\end{eqnarray}
where $f(k)$[$\zeta(k)$] is the radial part of the upper component
of the spinor $\psi(p)$[$F(p)$]. Similarly
$g(k)$[$\zeta^\prime(k)$] are the same of their lower component.
Functions $f(p)$ and $g(p)$ are the Fourier transforms of the radial
parts of the corresponding coordinate space spinors. $\zeta(k)$s are related
to $f$, $g$ and the scalar and vector potentials in the following way
\begin{eqnarray}
\zeta(k) & = & \zeta_s(k) - \zeta_v(k), \nonumber \\
\zeta^\prime(k) & = & \zeta^{\prime}_s(k) - \zeta^{\prime}_v(k),
\end{eqnarray}
where
\begin{eqnarray}
\zeta_s(k) & = & 4\pi \int dk^\prime k^{\prime 2} V_s(k^\prime)
             \frac{f(k+k^\prime)} {(1+k/k^\prime)^\ell},\\
\zeta_v(k) & = & 4\pi \int dk^\prime k^{\prime 2} V_v^0(k^\prime)
               \frac{f(k+k^\prime)} {(1+k/k^\prime)^\ell},\\
\zeta_s^\prime(k) & = & 4\pi \int dk^\prime k^{\prime 2} V_s(k^\prime)
               \frac{g(k+k^\prime)} {(1+k/k^\prime)^{\ell^\prime}},\\
\zeta_v^\prime(k) & = & 4\pi \int dk^\prime k^{\prime 2} V_v^0(k^\prime)
                \frac{g(k+k^\prime)} {(1+k/k^\prime)^{\ell^\prime}}.
\end{eqnarray}
More details of the derivations Eqs.~(21)-(28) can be found
in appendix A of the Ref.~\cite{shy95}. In Eq.~(23), 
${\mathscr Y}_{\ell 1/2 j}^{m_j}$ are the coupled spherical harmonics 
\begin{eqnarray}
{\mathscr Y}_{\ell 1/2 j}^{m_j} & = & <\ell m_\ell 1/2 \mu_i | j m_j>
                                Y_{\ell m_\ell}(\hat p) \chi_{1/2 \mu_i},
\end{eqnarray} 
where $\ell^\prime = 2\ell - j$ with $\ell$ and $j$ being the
orbital and total angular momenta, respectively and $Y$ represents the
spherical harmonics.  $\chi_{1/2 \mu_i}$ is the spin space
wave function of a spin-$\frac{1}{2}$ particle. In this paper we consider
only those cases where the final state has a pure single particle
configuration.  Likewise, the intermediate states are supposed to have
pure single-hole or particle structures.

The incident proton spinor and the outgoing kaon field are given by
\begin{eqnarray}
\Psi_i^{(+)}(p^\prime_i,p_i) & = & \delta(p^\prime_{i0} - E_i)
\sum_{J_pL_pM_p} <L_p M_p-\mu 1/2 \mu|J_p M_p> 
                    Y^*_{L_pM_p-\mu}(\hat{p_i}) \nonumber \\
  & \times &  {{F_{L_p J_p}(k^\prime_i,k_i) {\mathscr Y}_{L_p 1/2 J_p}^{M_p}
     ({\hat p}^\prime_i)} \choose {G_{L^\prime_p J_p}(k^\prime_i,k_i)
{\mathscr Y}_{L^\prime_p 1/2 J_p}^{M_p} ({\hat p}^\prime_i)}}, \\ 
\Phi_{K}^{(-)*}(p_K^\prime, p_K) & = &\delta(p^\prime_{K0}  - E_K) 
         \sum_{\ell_K m_K} (-)^{\ell_K} Y_{\ell_K m_K} ({\hat p}_K)
          Y_{\ell_K m_K}^*({\hat p}^\prime_K)\nonumber \\ 
         & \times &   f_{\ell_K}(k^\prime_K,k_K),
\end{eqnarray}
where $E_i$ and $E_K$ represents the energies of the incident proton and 
out going kaon, respectively, and  
where we have defined
\begin{eqnarray}
F_{L_p J_p}(k^\prime_i,k_i) & = & \frac{1}{2\pi^2} \int_0^\infty
       j_{L_p}(k^\prime_i a)\,F^C_{L_p J_p}(k_i,a) a^2 da, \\
G_{L^\prime_p J_p}(k^\prime_i,k_i) & = & \frac{1}{2\pi^2} \int_0^\infty
       j_{L^\prime_p}(k^\prime_i a)\,G^C_{L^\prime_p J_p}(k_i,a)
               a^2 da, \\
f_{\ell_K}(k^\prime_K,k_K) & = & \frac{1}{2\pi^2} \int_0^\infty
              j_{\ell_K}(k^\prime_K a)\,f^C_{\ell_k}(k_K,a) a^2 da.
\end{eqnarray}
In Eqs.~(32) and (33), $F^C_{L_p J_p}$ and $G^C_{L^\prime_p J_p}$
are the coordinate space spinors which are obtained by solving the
Dirac equation for the scattering state with appropriate complex scalar
and vector potentials. In Eqs.~(30)-(35), the momentum arguments $p$
denote the quantum numbers of the asymptotic free states whereas $p^\prime$
represent the momentum coordinates.
A short derivation of Eqs.~(30)-(31) is presented
in appendix B. In Eq.~(34) wave function
$f^C_{\ell_K}$ is the coordinate space solution of the Klein-Gordon
equation with kaon-nucleus optical potential (see, {\it e.g.},
Ref~\cite{tab77}). In these equations $j_\ell$ represent the spherical
Bessel function. It should be mentioned here that due to oscillatory nature
of the asymptotic these wave functions in the asymptotic region, the
integrals involved in Eqs.~(32)-(34) converge poorly. Such integrals can,
however, be calculated very accurately by using a contour integration
method~\cite{dav88}.

In the PW approximation, the wave functions $\Psi_i^{(+)}(p^\prime_i,p_i)$
and  $\Phi_{K}^{(-)*}(p_K^\prime, p_K)$ are given by
\begin{eqnarray}
\Psi_i^{(+)}(p^\prime_i,p_i) & = &\sqrt{\frac{E_i + m_N}{2m_N}} {{\chi_{\mu_i}}
                              \choose
              \frac{\sigma \cdot p}{E_i + m_N} \chi_{\mu_i}}
              \delta^4(p^\prime_i - p_i),\\
\Phi_{K}^{(-)*}(p_K^\prime, p_K) & = & \delta^4(p^\prime_K - p_K).
\end{eqnarray}
In Eq.~(35) $\chi_{\mu_i}$ represents the two component Pauli spinor.
The other symbols have the same meaning as in Ref.~\cite{bjo64}.

By repeated use of Eqs.~(21) and (22) one can perform the tedious but
straight forward algebra to trace out the $\gamma$ matrices from the 
expression for the amplitudes [e.g., Eq.~(16)]. Furthermore, by using 
Eqs. (21)-(28) and (30)-(33) and performing the angular momentum algebra
one can get the amplitudes in a form which is suitable for numerical
evaluation. We write in appendix A, {\it e.g.,} the final form of the
amplitude $M_{2b}$ [Eq.~(16)] in both full distorted wave as well
as plane wave approximations.

To get the $T$ matrix of the $(p,K^+)$ reaction, one has to integrate
the amplitude corresponding to each graph over all the independent
intermediate momenta subject to constraints imposed by the
momentum conservation at each vertex. For instance, for the amplitude
corresponding to Eq.~(16) the respective $T$ matrix is given by
\begin{eqnarray}
T_{2b}(N^*_{1/2}) & = & \int \frac{d^4p^\prime_i}{(2\pi)^4}
\int \frac{d^4p^\prime_K}{(2\pi)^4}
\int \frac{d^4p_\Lambda}{(2\pi)^4} \int
\frac{d^4p_2}{(2\pi)^4} \nonumber \\
& \times &
\delta(q - p_1 + p_2)\delta(p_{N^*} - p_{K} - p_\Lambda)\nonumber \\
& \times &
\delta(p_\Lambda - p_i^\prime - q + p_K^\prime) M_{2b}(N^*_{1/2}).
\end{eqnarray}
It can be seen, from Eq.~(37), that in the full distorted wave theory
one would be required to perform a twelve dimensional integration to evaluate
the $T$-matrix $T_{2b}$ which requires a tremendous numerical effort.
This gets further complicated due to the fact that the number of 
partial waves required in Eqs.~(30)-(32) are quite large for these
high energy reactions. On the other hand, in the plane wave approximation
the integrations over variables $p_i^\prime$ and $p_K^\prime$ become
redundant. This not only reduces the dimensionality of the integrations
by a factor of 2 but also removes the requirement of partial wave 
summations altogether. In this exploratory study we would like to
make calculations for very many cases involving a variety of targets and
beam energies in order to understand the basic mechanism of this reaction
and to get the estimates of the order of magnitudes of various cross
sections to guide the planning of the future experiments. We have,
therefore, made our calculations in a numerically simpler plane wave
theory. However, estimates of the distortions have been made within
the eikonal approximation.

The differential cross section for the $(p,K^+)$ reaction is given by
\begin{eqnarray}
\frac{d\sigma}{d\Omega} & = & \frac{1}{(4\pi)^2}
\frac{m_pm_Am_B}{(E_{p_i} + E_A)^2} \frac{p_K}{p_i}
 \sum_{m_im_f} |T_{m_im_f}|^2,
\end{eqnarray}
where $E_{p_i}$ and $E_A$ are the total energies of incident
proton and the target nucleus, respectively while $m_p$, $m_A$ and $m_B$ are
the masses of the proton, and the target and residual nuclei, respectively.
The summation is carried out over initial ($m_i$) and final ($m_f$)
spin states.  $T$ is the final $T$ matrix obtained by summing the transition
matrices corresponding to all the graphs.

\section{Results and Discussions}
\subsection{The nuclear and hypernuclear structure and bound state spinors}
\begin{figure}
\begin{center}
\mbox{\epsfig{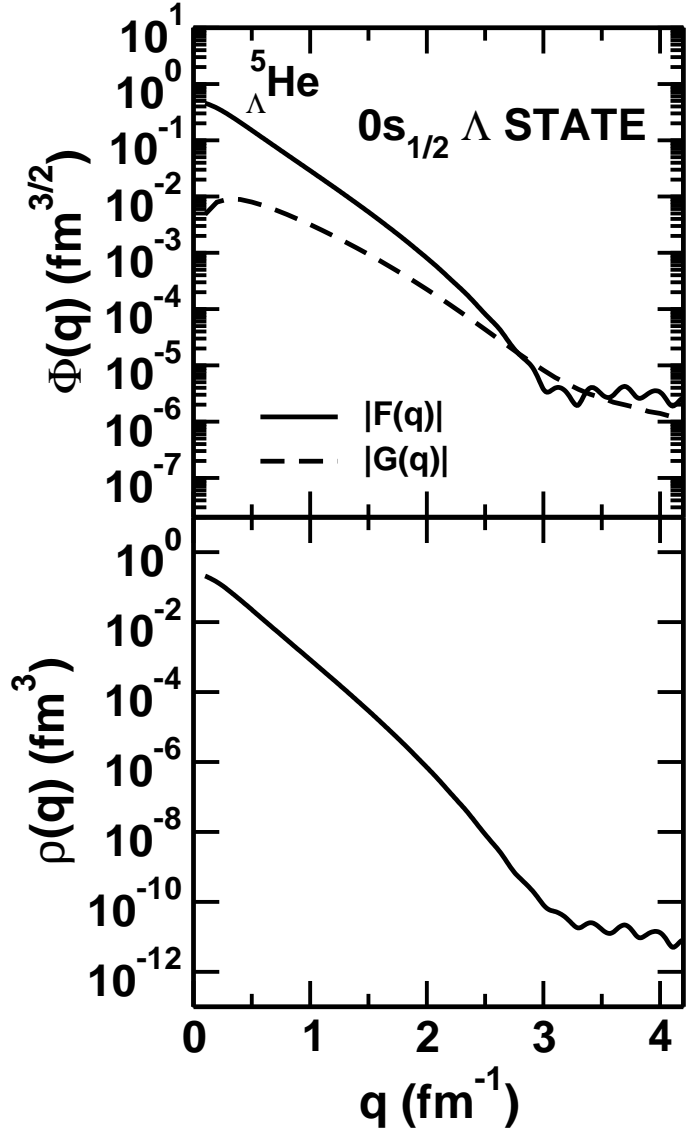}}
\end{center}
\caption{(Upper panel) Momentum space spinors [$\Phi(q)$] for
$0s_{1/2}$ $\Lambda$ orbit in $^{5}\!\!\!_\Lambda$He hypernucleus.
$|F(q)|$ and $|G(q)|$ are the upper and lower components
of the spinor, respectively. (Lower panel) Momentum distribution 
[$\rho(q) = |F(q)|^2 + |G(q)|^2$] for
the same hyperon state calculated with these wave functions.  
}
\label{fig:figd)}
\end{figure}

The spinors for the final bound hypernuclear state (corresponding to
momentum $p_\Lambda$) and for two intermediate nucleonic states
(corresponding to momenta $p_1$ and $p_2$) are required to 
perform numerical calculations of various amplitudes. We assume these
states to be of pure-single particle or single-hole configurations 
with the core remaining inert. To simplify the nuclear structure problem
the quantum numbers of the two intermediate states are taken to be the
same although it is straight forward to include also those cases where
they may occupy different orbits. The spinors in the momentum space are
obtained by Fourier transformation of the corresponding coordinate space
spinors which are the solutions of the Dirac equation with potential fields
consisting of an attractive scalar part ($V_s$) and a repulsive vector part
($V_v$) having a Woods-Saxon form. This choice appears justified as the
Dirac Hartree-Fock calculations~\cite{mil72,bro78} suggest that these
potentials tend to follow the nuclear shape. The same potential form has
also been used in the relativistic ONM~\cite{coo82} and TNM
calculations~\cite{shy95} of the $(p,\pi)$ reaction.
\begin{figure}
\begin{center}
\mbox{\epsfig{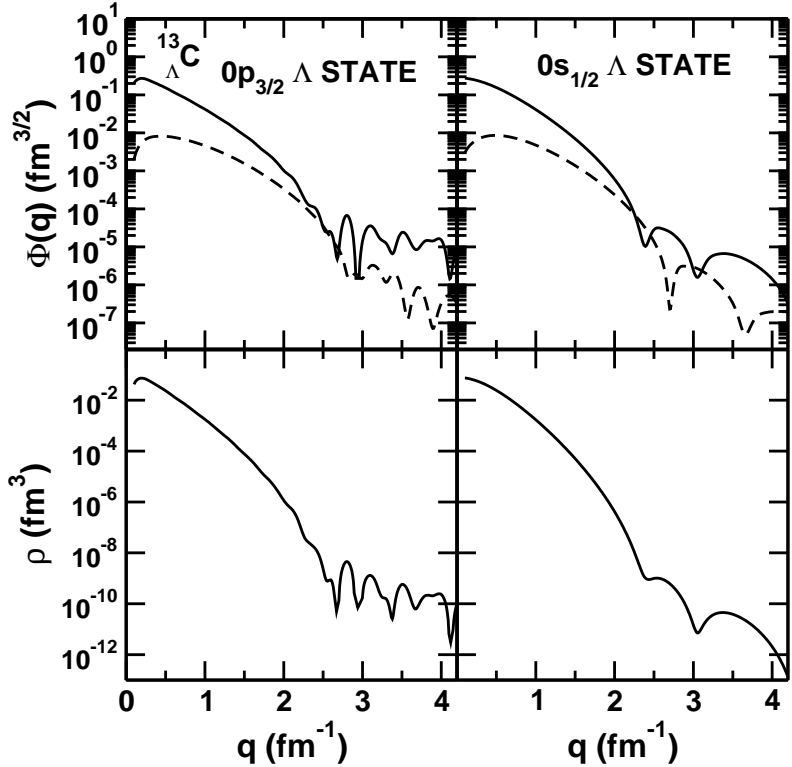}}
\end{center}
\caption{same as in Fig.~4 for $0p_{3/2}$ $\Lambda$ and
$0s_{1/2}$ $\Lambda$ orbits in $^{13}\!\!\!_\Lambda$C hypernucleus.
}
\label{fig:fige)}
\end{figure}

With a fixed set of the geometry parameters (reduced radii
$r_s$ and $r_v$ and diffusenesses $a_s$ and $a_v$), the depths of the
potentials were searched in order to reproduce the binding energies of
the particular state (the corresponding values are given in Table 2).
We use the same geometry for the scalar and vector potentials. The
depths of potentials $V_s$ and $V_v$ were further constrained by
requiring that their ratios are equal to -0.81 as suggested in 
Ref.~\cite{ser86}. Experimental inputs have been used for the 
configurations and the single baryon binding energies of the states in
hypernuclei $^5$$\!\!\!_\Lambda$He~\cite{ban90} and 
$^{13}$$\!\!\!_\Lambda$C~\cite{ban90,may81}. However, the quantum numbers
as well as the single baryon binding energies for states in
$^{41}$$\!\!\!_\Lambda$C hypernucleus have been taken from the density
dependent relativistic hadron field (DDRH) theory predictions of
Ref.~\cite{kei00}, which reproduce the corresponding experimental binding
energies reasonably well. In this case, we have compared, for each state,
the spinors calculated by our well-depth search method with those
calculated within the DDRH~\cite{kei00} theory and find an excellent
agreement between the two. 
\begin{table}
\begin{center}
\caption[T2] {Searched depths of vector and scalar potentials and the binding
energies of the $\Lambda$ and nucleon bound states. 
}
\vspace{1.0cm}
\begin{tabular}{|c|c|c|c|c|c|c|c|} \hline
 State & Binding Energy ($\epsilon$)& $V_v$ & $r_v$ & $a_v$ & $V_s$ & $r_s$ & $a_s$\\
       &(\footnotesize{MeV})&(\footnotesize{MeV})&(\footnotesize{fm})&
  (\footnotesize{fm}) & (\footnotesize{MeV}) & (\footnotesize{fm}) &
(\footnotesize{fm}) \\   
\hline
$^{5}\!\!\!_\Lambda$He$(0s_{1/2})$ & 3.12 & 135.424 & 0.983 & 0.578 &
-167.190 & 0.983 & 0.606  \\
$^{4}$He$(0s_{1/2})$ & 19.814 & 384.053 & 0.983 & 0.578 & -429.695 &
0.983 & 0.606  \\
$^{13}\!\!\!_\Lambda$C$(0p_{3/2})$ & 0.860 & 180.712 & 0.983 & 0.578 &
-223.101 & 0.983 & 0.606  \\
$^{13}\!\!\!_\Lambda$C$(0p_{1/2})$ & 0.708 & 206.161 & 0.983 & 0.578 &
-254.520 & 0.983 & 0.606  \\
$^{13}\!\!\!_\Lambda$C$(0s_{1/2})$ & 11.690 & 166.473 & 0.983 & 0.578 &
-205.522 & 0.983 & 0.606  \\
$^{12}$C$(0p_{3/2})$ & 15.957 & 382.598 & 0.983 & 0.578 & -472.343 &
0.983 & 0.606  \\
$^{41}\!\!\!_\Lambda$Ca$(0s_{1/2})$ & 17.882 & 154.884 & 0.987 & 0.676 & 
-191.215 & 0.982 & 0.700  \\
$^{41}\!\!\!_\Lambda$Ca$(0p_{3/2})$ & 9.677  & 179.485 & 0.987 & 0.676 &
-221.587 & 0.982 & 0.700 \\
$^{41}\!\!\!_\Lambda$Ca$(0p_{1/2})$ & 9.140  & 188.188 & 0.987 & 0.676 &
-232.331 & 0.982 & 0.700 \\
$^{41}\!\!\!_\Lambda$Ca$(0d_{5/2})$ & 1.544  & 207.490 & 0.987 & 0.676 &
-256.160 & 0.982 & 0.700 \\
$^{41}\!\!\!_\Lambda$Ca$(1s_{1/2})$ & 1.108  & 192.044 & 0.987 & 0.676 &
-237.091 & 0.982 & 0.700 \\
$^{41}\!\!\!_\Lambda$Ca$(0d_{3/2})$ & 0.753  & 230.206 & 0.987 & 0.676 &
-284.205 & 0.982 & 0.700  \\
$^{40}$Ca$(0d_{3/2})$ & 8.333  & 360.980 & 0.987 & 0.676 & -445.660 &
0.982 & 0.700 \\
\hline
\end{tabular}
\end{center}
\end{table}

In Figs.~(4)-(6) we show the lower and upper components of the Dirac 
spinors in momentum space for the $0s_{1/2}$ hyperon in 
$^{5}\!\!\!_\Lambda$He, $0p_{3/2}$ and $0s_{1/2}$ hyperons in 
$^{13}\!\!\!_\Lambda$C, and $0d_{3/2}$ and $0p_{1/2}$ hyperons in
$^{41}\!\!\!_\Lambda$Ca, respectively. In each case, we note that only
for momenta $<$ 1.5 fm$^{-1}$, is the lower component of
the spinor substantially smaller than the upper component. In the region
of momentum transfer pertinent to exclusive kaon production in 
proton-nucleus collisions, the lower components of the spinors are not
negligible as compared to the upper component which clearly demonstrates
that a fully relativistic approach is essential for an accurate
description of this reaction.
 
The spinors calculated in this way provide a good description of the
experimental nucleon momentum distributions for various nucleon orbits
as is shown in Ref.~\cite{shy95}. In the lower panel of each of
Figs.~(4)-(6), we show momentum distribution $\rho(q)\,[ = 
F(q)|^2 + |G(q)|^2]$~\cite{fru84} of the corresponding $\Lambda$ hyperon.
It can be noted that in each case the momentum density of the hyperon shell,
in the momentum region around 0.35 GeV/c, is at least 2-3 orders of magnitude
larger that around 1.0 GeV/c. Since, in the TNM the large momentum is shared
among three baryons, the model is sensitive to the bound state spinors in the 
momentum regime where they are well described and are quite large.
Therefore, one expects to get a larger cross section for exclusive
meson production reactions within this model. 
\begin{figure}
\begin{center}
\mbox{\epsfig{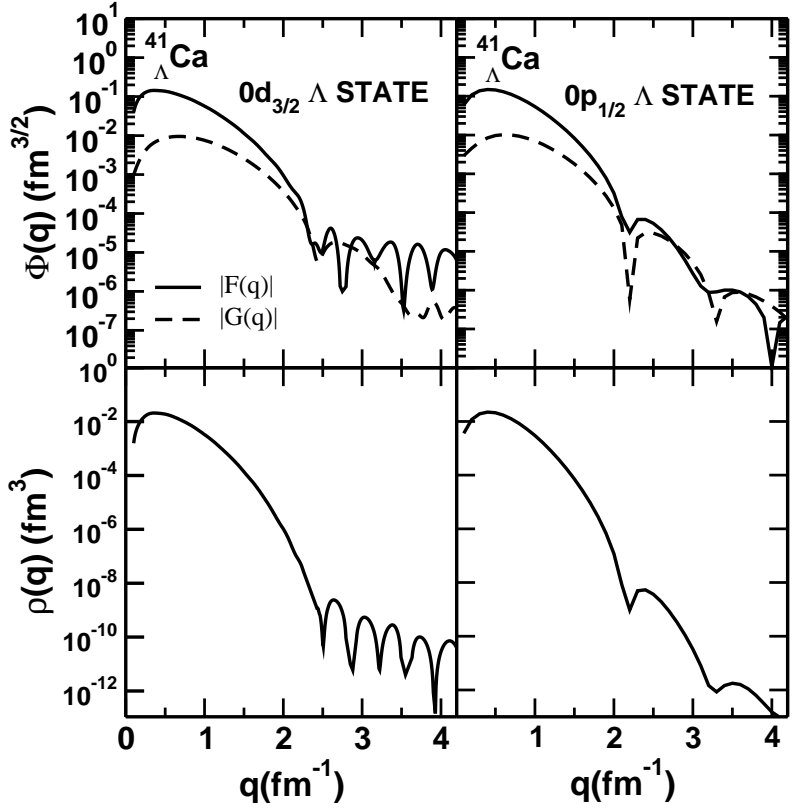}}
\end{center}
\caption{Same as in Fig.~4 for $0d_{3/2}$ $\Lambda$
and $0p_{1/2}$ $\Lambda$ orbits in $^{41}\!\!\!_\Lambda$Ca hypernucleus.
}
\label{fig:figf)}
\end{figure}

\subsection{Cross sections for A($p,K^+)_\Lambda$B reaction} 

\subsubsection{Results within the plane wave approximation}

The self-energies $\Pi_\pi$, $\Pi_\rho$ and $\Pi_\omega$ of pion, rho
and omega mesons, respectively, are another input quantities required
in the calculations of the kaon production amplitudes. They take into
account the medium effects on the intermediate meson propagation. The 
$\rho$ and $\omega$ self energies have been calculated by following
the procedure described in Ref.~\cite{muh04}. It may be noted that
the intermediate mesons are always space-like and their self energies
are functions of the exchanged four momenta.

The pion self energy, which is more crucial as one-pion exchange
diagrams dominate the $(p,K^+)$ cross sections, is the same as that shown
in Ref.~\cite{shy04}. $\Pi_\pi$ is obtained by calculating the contribution
of particle-hole ($ph$) and delta-hole ($\Delta h$) excitations
produced by propagating pions~\cite{dmi85}. This has been renormalized by
including the short-range repulsion effects by introducing the constant
Landau-Migdal parameter $g^\prime$ which is taken to be the same for
$ph-ph$ and $\Delta h-ph$ and $\Delta h-\Delta h$ correlations which is
a common choice. The parameter $g^\prime$, acting in the spin-isospin 
channel, is supposed to mock up the complicated density dependent effective
interaction between particles and holes in the nuclear medium. Most
estimates give a value of $g^\prime$ between 0.5 - 0.7. The sensitivity of
the $(p,K^+)$ cross sections to the parameter $g^\prime$ is discussed in
Ref~\cite{shy04}  where a detailed discussion is presented of the effect of the
self-energy correction (to the pion propagator) on various amplitudes.
\begin{figure}
\begin{center}
\mbox{\epsfig{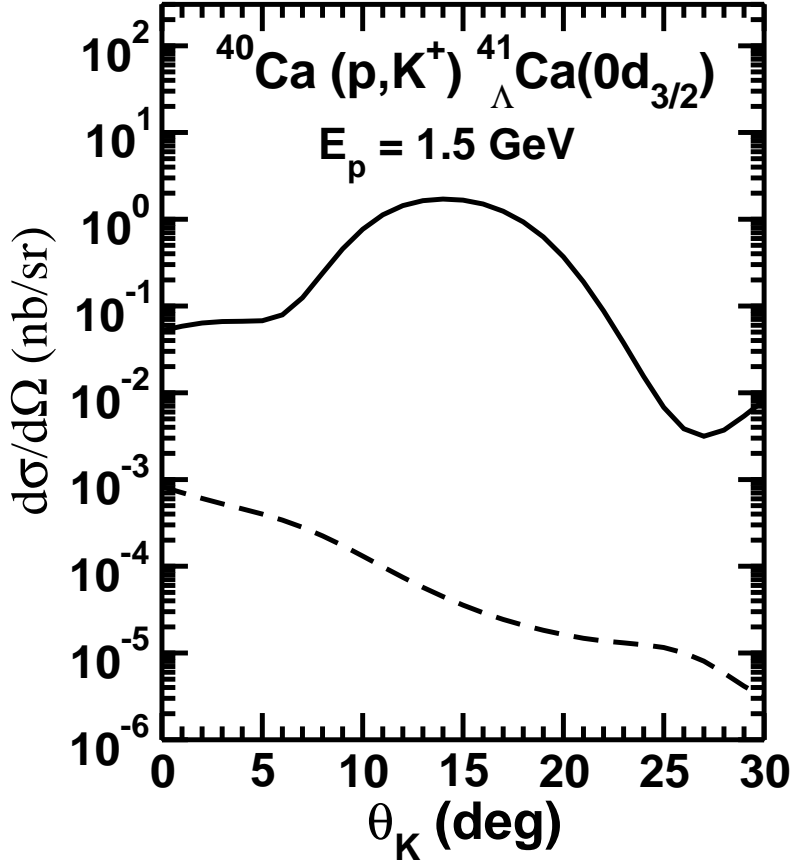}}
\end{center}
\caption{Relative contributions of target emission (TE) (solid line) 
and projectile emission (PE) (dashed line) graphs to the differential
cross section for the
$^{40}$Ca$(p,K^+)$$^{41}\!\!\!_\Lambda$Ca$(0d_{3/2})$ reaction 
at the beam energy of 1.5 GeV. Only pion-exchange graphs are included
in these calculations. The pion self-energy used in these calculations
has been renormalized by using the Landau parameter $g^\prime = 0.5$. 
}
\label{fig:figg)}
\end{figure}

In the PE graph the values of the time-like momentum component $q_0$
in the meson propagator is taken to be $q_0 = E_i$, where the latter
is the energy of the incident proton (ignoring the nucleon binding
energies) by the energy conservation at the vertices. Due to this 
the meson propagator in this diagrams [Fig.~2(c)] may develop a pole in
the integration at $|q| = \sqrt{|q|^2+m_\lambda^2}$ ($m_\lambda$ is the
mass of the meson).  With the inclusion of the self-energy (which, in general,
is complex at non-zero energies), the pole is automatically removed from
the real axis and the pole integration problem disappears and the graph
2(c) becomes well behaved. Furthermore, this leads to a strong reduction
in the contributions of this graph.
\begin{figure}
\begin{center}
\mbox{\epsfig{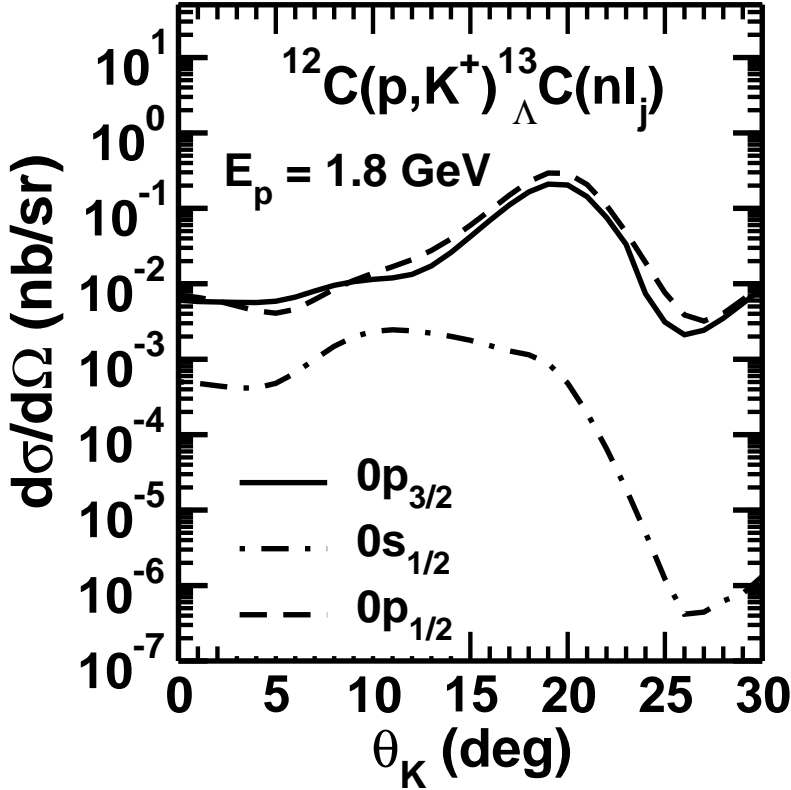}}
\end{center}
\caption{Differential cross section for the 
$^{12}$C$(p,K^+)$$^{13}\!\!\!_\Lambda$C
reaction for the incident proton energy of 1.8 GeV for various
bound states of final hypernucleus as indicated in the figure.
The $\Lambda$ separation energies for various states were the same
as those shown in Table 2. 
}
\label{fig:figh)}
\end{figure}

On the other hand, such a pole does not develop in the TE diagram [Fig.~2(b)]
since in this case the pion self-energy is real and attractive
due to $q_0 \approx 0$ [as can be seen from Fig.~3 of Ref.~\cite{shy04}].
This leads to a small denominator of the pion propagator leading to a
consequent increase in the contribution of this graph. In fact this
amplitude dominates the $(p,K^+)$ cross sections as can be seen from
Fig.~7 where we show the relative contributions of graph 2(b) and 2(c) to
this cross section considering only the one pion exchange mechanism.
Similar domination of the TE amplitudes have also been seen in case of the
TNM calculations of the ($p,\pi$) reaction~\cite{shy95}. 
        
In Figs.~8 and 9, we show the kaon angular distributions corresponding
to various final hypernuclear states excited in reactions
$^{12}$C$(p,K^+)$$^{13}\!\!\!_\Lambda$C and \\
$^{40}$Ca$(p,K^+)$$^{41}\!\!\!_\Lambda$Ca, respectively. The incident  
proton energies in the two cases are taken to be 1.8 GeV and 1.5 GeV,
respectively where the angle integrated cross sections for the two
reactions are maximum. In both the cases we have taken the Landau
parameter $g^\prime = 0.5$ in the calculations of the pion self energy.
The calculations are the coherent sum of all the amplitudes corresponding
to the various meson exchange processes and intermediate resonant states.
Clearly, the cross sections are quite selective about the excited
hypernuclear state, being maximum for the state of largest orbital
angular momentum. This is due to the large momentum transfer involved
in this reaction. It may be noted that in each case the angular
distribution has a maximum at angles larger than 0$^\circ$.  
This is  due to the fact that there are several maxima in the upper
and lower components of the momentum space bound spinors in the
region of large momentum transfers. Therefore, in the kaon angular
distribution the first maximum may shift to larger angles
reflecting the fact that the bound state wave functions show
diffractive structure at higher momentum transfers.

We see that while in Fig.~(8) there is a difference of more than
an order of magnitude between the cross sections for the $p$-shell and 
the $s$-shell excitations, the two are almost of the same 
order of magnitude in Fig.~9. To understand this, we note from Table 2
that for the $^{13}\!\!\!_\Lambda$C hypernucleus, the binding energies of the
$p$-shell states are in the range of only 0.70-0.90 MeV as compared to
11.69 MeV of the $s$-shell state. The bound state spinor ($\Phi$)
behaves, at larger momentum transfers, approximately as $\frac{1}{k^2}$
where $k^2 \propto \epsilon$. Thus, with increasing binding energies
the magnitudes of $\Phi$ decreases at larger momenta. This leads to a
decrease in the cross sections of those processes which are more 
sensitive to the larger momentum transfers. The $p$-shell state spinors
in $^{13}\!\!\!_\Lambda$C, are considerably larger than those of the $s$
state in the relevant momentum transfer ($q$) region ($\sim$ 2-4 fm$^{-1}$).
Therefore, the $p$-shell transitions are enhanced as compared to the
$s$-shell one. Due to this binding energy selectivity, we call the 
reaction leading to the $p$-shell excitation as a matched transition while
that to the $s$-shell a mismatched one. 

On the other hand, both $p$- and $s$-shell transitions in Fig.~9 are
strongly mismatched due to the fact that for the $^{41}\!\!\!_\Lambda$Ca
hypernucleus both these states have large binding energies as compared to
those of the $d$-shell. In this case, the spinors of the $p$- and $s$-states
do not differ substantially from each other in the relevant $q$ region,
which in turn makes the corresponding cross sections not to differ too
much from each other. As a matter of fact, states with larger angular
momenta are squeezed in the coordinate space (r-space) due to the centrifugal
barrier and hence they are spread out in the momentum space (q-space) which 
enhances the cross sections for correponding transitions. It should,
however, be mentioned here that for a {\it fixed angular momentum}, states
with larger binding are more compact in r-space, and therefore are wider in
q-space. Hence, in such cases cross sections to states with larger
binding may actually be larger. 
\begin{figure}
\begin{center}
\mbox{\epsfig{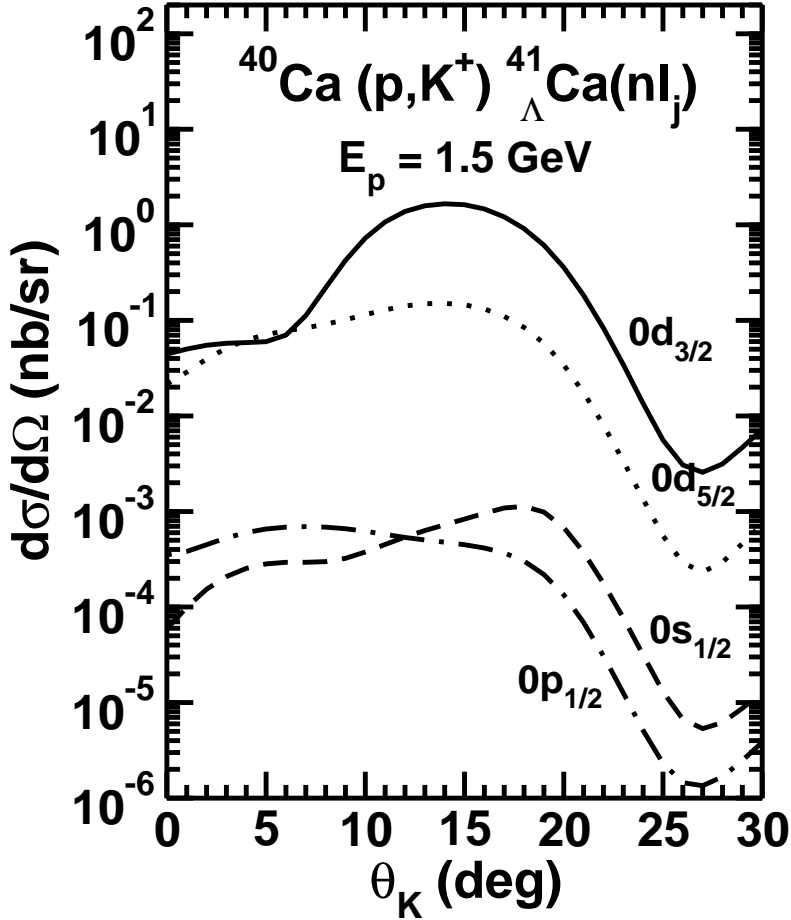}}
\end{center}
\caption{Differential cross section for the 
$^{40}$Ca$(p,K^+)$$^{41}\!\!\!_\Lambda$Ca reaction for the incident
proton energy of 1.5 GeV for various bound states of final hypernucleus
as indicated in the figure. The $\Lambda$ separation energies for
various states were as shown in Table 2.
}
\label{fig:figi)}
\end{figure}

The absolute magnitudes of the cross sections near the peak is around
1-2 nb/sr, although the distortion effects could reduce these values 
as is shown below. This order of magnitude estimates should be
useful in planning of the future experiments for this reaction. As found
in Ref.~\cite{shy04} contributions from the $N^*$(1710) resonance dominate
the total cross section in each case. We also note that the interference
terms of the amplitudes corresponding to various resonances are not
negligible. It should be emphasized that we have no freedom in choosing
the relative signs of the interference terms.
\begin{figure}
\begin{center}
\mbox{\epsfig{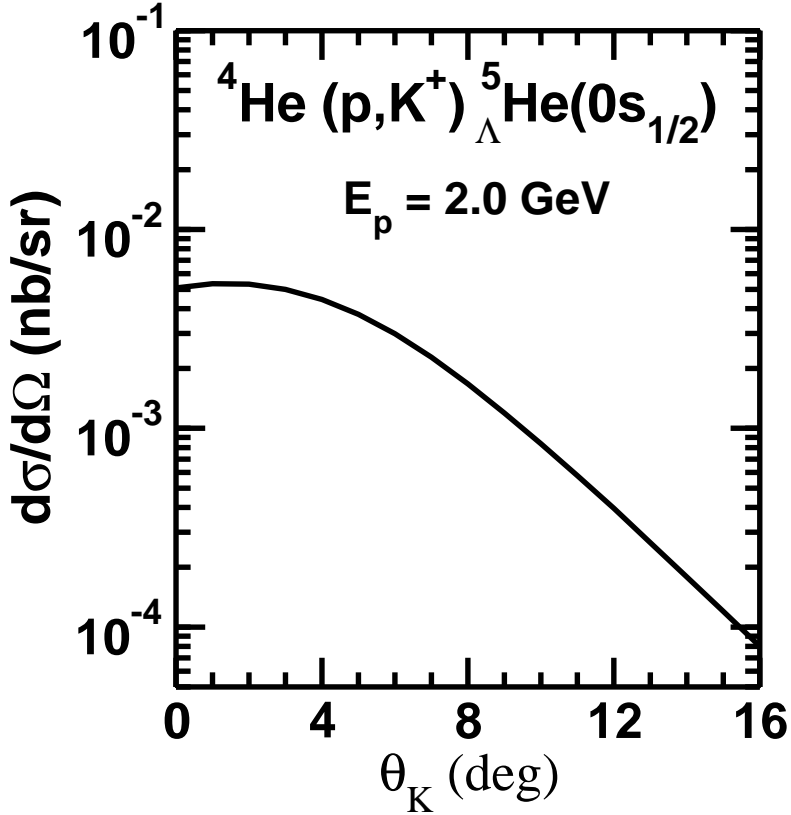}}
\end{center}
\caption{Differential cross section for the 
$^{4}$He$(p,K^+)$$^{5}\!\!\!_\Lambda$He
reaction for the incident proton energy of 2.0 GeV for the 
bound state of final hypernucleus as indicated in the figure.
The $\Lambda$ separation energy was as shown in Table 2.
}
\label{fig:figj)}
\end{figure}

In Fig.~(10), we show the kaon angular distributions for 
reaction $^{4}$He$(p,K^+)$$^{5}\!\!\!_\Lambda$He leading to the
$0s_{1/2}$ state of the hypernucleus at the proton beam energy of 2.0
GeV. In contrast to Figs~8 and 9, the maximum in the cross section,
in this case, occurs at zero degree and it decreases gradually as the
angle increases.
This difference in the pattern of angular distribution seen in Fig.~10
from that of Figs.~8 and 9 can be understood from the fact that 
for momentum transfers relevant to this case the Dirac spinors are
smoothly varying and are devoid of structures as can be seen in Fig.~4.
In any case, from the purely kinematical arguments it is clear that the
maximum in the cross section for a $\ell = 0$ transition is expected
normally to occur at smaller angles as compared that for a $\ell \neq 0$
one. The absolute magnitude of the cross section at the forward angles, in
this case, is about 0.02 nb/sr.  This value is somewhat smaller than 
the upper limit of the experimental center of mass cross section deduced
for this reaction in a very preliminary study made in Ref.~\cite{kin98} at 
a similar beam energy~\cite{kin98}.
\begin{figure}
\begin{center}
\mbox{\epsfig{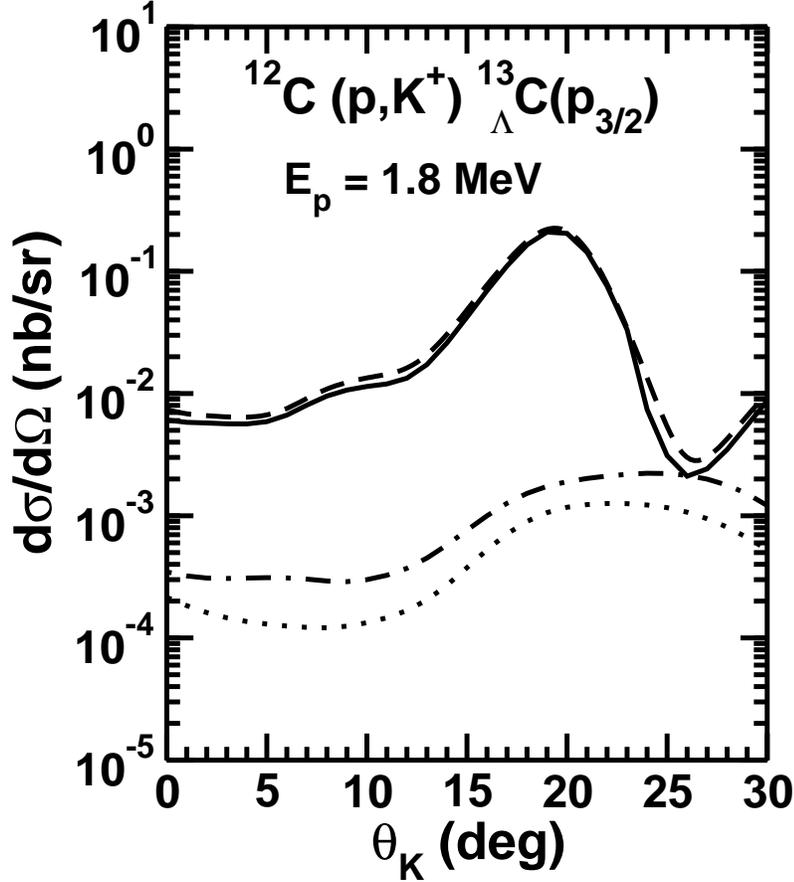}}
\end{center}
\caption{Differential cross section for the 
$^{12}$C$(p,K^+)$$^{13}\!\!\!_\Lambda$C$(0p_{3/2})$
reaction for the incident proton energy of 1.8 GeV. 
The dashed, dashed-dotted
and dotted line show the contributions of $\pi$, $\rho$ and $\omega$
exchange processes, respectively. Their coherent sum is depicted by
the full line.  }
\label{fig:figk)}
\end{figure}

In Fig.~(11)-(12), we investigate the role of various meson exchange 
processes in describing the differential cross sections as a function
of outgoing kaon angles for reactions
$^{12}$C$(p,K^+)$$^{13}\!\!\!_\Lambda$C$(0p_{3/2})$, and 
$^{40}$Ca$(p,K^+)$$^{41}\!\!\!_\Lambda$Ca$(0d_{3/2})$ at beam energies of
1.8 GeV, and 1.5 GeV, respectively. In each case, dashed,
dashed-dotted and dotted lines show the contributions of $\pi$, $\rho$
and $\omega$ exchange processes, respectively. Their coherent sum is
depicted by the full line. We note that pion exchange graphs
dominate the production process for all angles in each case. 
This result is similar to the observation made in the case of
associated kaon production in elementary $pp$ collisions~\cite{shy99,fae97}. 
For the $^{12}$C$(p,K^+)$$^{13}\!\!\!_\Lambda$C$(0p_{3/2})$, and 
$^{40}$Ca$(p,K^+)$$^{41}\!\!\!_\Lambda$Ca$(0d_{3/2})$ reactions,
contributions of $\rho$ and $\omega$ exchange processes are 
unimportant at forward angles. However, at backward angles they become
relatively more significant. This is understandable as at the backward angles
the momentum transfer is larger which leads to greater contributions from
the heavier meson exchanges. We see a destructive interference 
between the $\pi$ and $\rho$ contributions.
The $\rho$ and $\omega$ exchange contributions put together do not affect
appreciably the pion exchange only cross sections even at the backward
angles due their mutual cancellation.
\begin{figure}
\begin{center}
\mbox{\epsfig{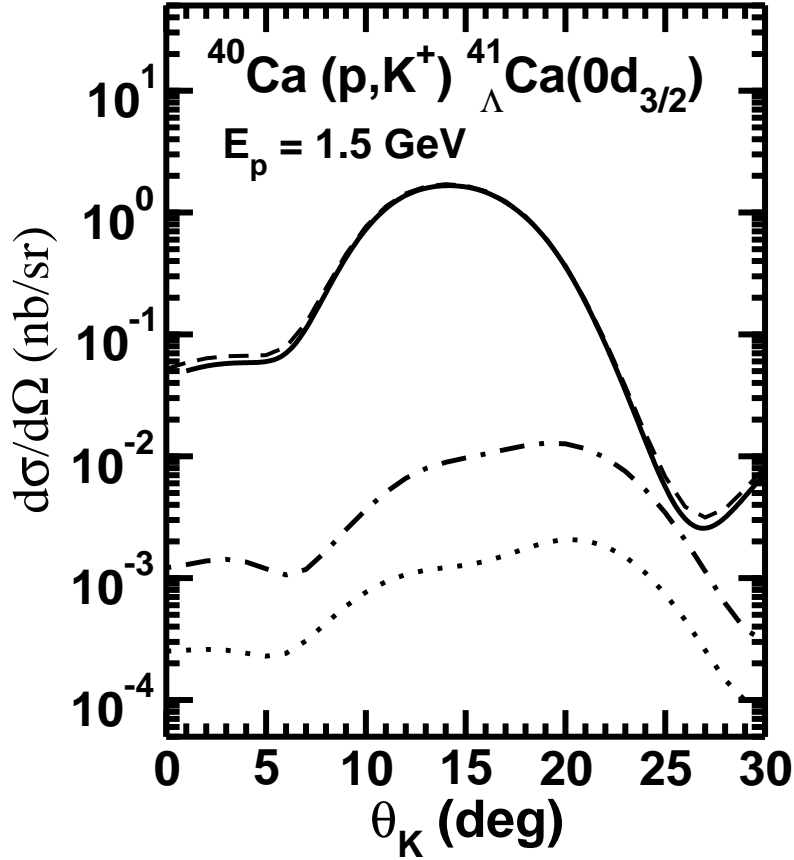}}
\end{center}
\caption{Differential cross section for the 
$^{40}$Ca$(p,K^+)$$^{41}\!\!\!_\Lambda$Ca$(0d_{3/2})$
reaction for the incident proton energy of 1.5 GeV. various curves have
the same meaning as in Fig.~11.
}
\label{fig:figl)}
\end{figure}

In Figs.~13 and 14 we show the predicted beam energy dependence of the 
angle integrated cross sections for 
$^{12}$C$(p,K^+)$$^{13}\!\!\!_\Lambda$C$(0p_{3/2})$ and
$^{40}$Ca$(p,K^+)$$^{41}\!\!\!_\Lambda$Ca$(0d_{3/2})$
reactions, respectively. The results indicate a substantial dependence
on the beam energy of the integrated cross sections. As the beam 
energy increases beyond the respective threshold energies, $E_p^{thres}$,
the cross section first increases fast and then after peaking around a 
given value starts decreasing slowly. This trend is reminiscent of the
general nature of the beam energy dependence of associate kaon production
cross sections in elementary $pp$ collisions. The slight shift in the
peak position towards the smaller value in the case of the heavier
nucleus reflects the fact that the value of $E_p^{thres}$ is smaller as
one goes to heavier target. 
 
\subsubsection{Effects of initial and final state interactions }

Calculations presented thus far have been done in a plane wave 
approximation, where the proton-nucleus and kaon-nucleus interactions
are ignored. While the essential features of the high momentum
transfer reaction $(p,K^+)$ can be understood in this approach,
the nuclear interactions may have some consequences. They produce 
both absorptive and dispersive effects~\cite{dov80}. However, for
large incident energies considered in this calculation, the
absorption effects are likely to be the most important; a qualitative
estimate of this is provided in the following.
\begin{figure}
\begin{center}
\mbox{\epsfig{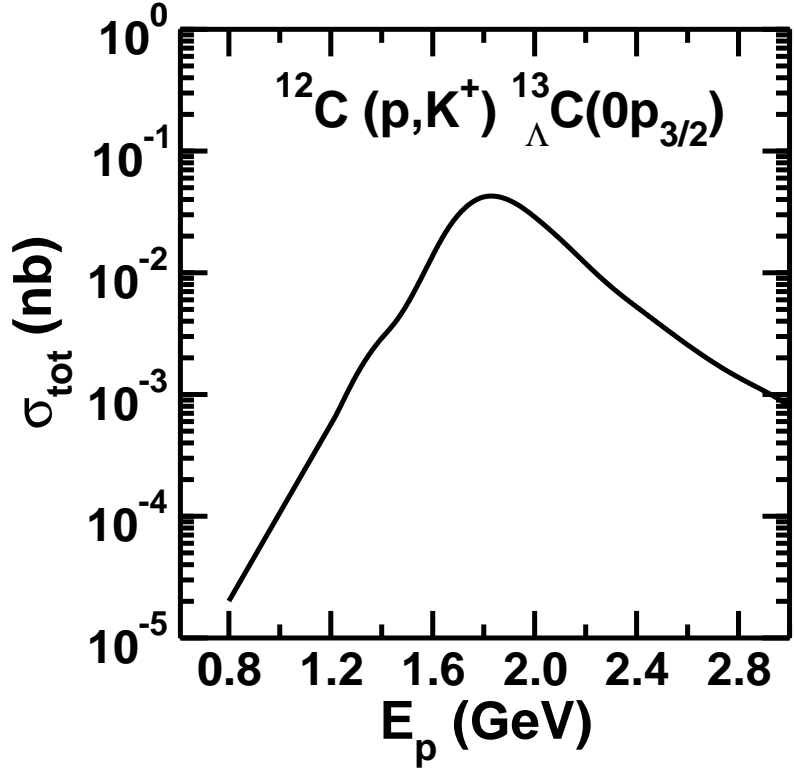}}
\end{center}
\caption{Angle integrated cross section for the 
$^{12}$C$(p,K^+)$$^{13}\!\!\!_\Lambda$C$(0p_{3/2})$ reaction 
as a function of the incident proton beam energy. 
}
\label{fig:fign)}
\end{figure}

We use the eikonal approximation to estimate the attenuation factor
for a particle traveling through the nuclear medium. First, we
define a refractive index of the nuclear medium as (see, e.g.,
\cite{wei88})
\begin{eqnarray} 
\eta(r,E) & = & \eta(E) \frac{\rho(r)}{\rho_0},
\end{eqnarray}
where $\rho(r)[\rho(\sqrt{b^2+z^2}]$ is the nuclear density distribution,
and
\begin{eqnarray}
\eta(E) & = & \frac{\kappa(E)}{k(E)}.
\end{eqnarray}
Here $k$ is the external wave number, and $\kappa$ is the wave
number in the medium. The imaginary part of this refractive index is
defined as $\eta_0$. In terms of $\eta_0$, the attenuation factor can be 
written as~\cite{jai88}
\begin{eqnarray}
S(E) & = & \int\, d{\bf b}\,dz\, \frac{\rho({\bf b},z)exp[-k\,\eta_0(E)
         L(b)]}{\int\,d{\bf b}\,dz\,\rho({\bf b},z)},
\end{eqnarray}
where $L(b)$ is the length of the path traveled by the particle in the
medium, which is given by
\begin{eqnarray}
L(b) & = & \int_0^\infty \frac{\rho(r)}{\rho_0}\,dz
\end{eqnarray}
If the nuclear density is approximated by a Gaussian function,\\
$\rho(r) = \rho_0\,exp\,(-r^2/\alpha^2)$, the integration in Eq.~(41)
can be done analytically.        
In this case the attenuation factor is given by
\begin{eqnarray}
S(E) & = & \frac{1\,-\,exp[\,-\,\sqrt{\pi}\,\alpha\,k\,\eta_0(E)]}{
         \sqrt{\pi}\,k\,\eta_0(E)}
\end{eqnarray}
The attenuation due to medium can be calculated if the value of 
$\eta_0(E)$ is known. This can be obtained from the imaginary part
of the optical potential $W_0$ as
\begin{eqnarray}
\eta_0(E) & = & \frac{1}{\hbar^2}\frac{E}{k^2}W_0\,(E)
\end{eqnarray}
We use here the following high energy relations to obtain $W_0$
\begin{eqnarray}
W_0\,(E) & = & \hbar^2 \frac{k\,\sigma_T \rho_0}{2E},
\end{eqnarray} 
where $\sigma_T$ is the total nucleon-nucleon or meson-nucleon cross
section. In order to determine the total attenuation factor for the $(p,K^+)$
reaction, the total attenuation due to both the proton and kaon 
distortions has been estimated by replacing the factor $k\eta_0$ in
Eq.~(43) by
\begin{eqnarray}
k\eta_0 &  \rightarrow & k_p \eta_0(E_p) + k_K \eta_0(E_K)
\end{eqnarray}
In our estimation of the attenuation we have taken the average values of
$\sigma_T$ as 45 mb and 14 mb for $pp$~\cite{bug66} and $K^+p$ systems
\cite{sah04}, respectively. The value of the parameter $\alpha$ is taken
to be 1.61 fm, 2.37 fm and 3.52 fm for $^4$He, $^{12}$C and $^{40}$Ca
targets, respectively~\cite{hof56}. In Table 3, we show the values of 
the total attenuation $|S(E)|^2$ for the three targets discussed
in this paper. We see that overall effect of initial and final state
interactions is to decrease the peak cross sections relative to the PW
calculations by a factor of 3 - 10. We however, would like to emphasize that
these estimates only provide an idea of the absorptive effects.
In more rigorous calculations, refractive distortion effects also play
a role which could influence the shapes of the angular distributions as well. 
\begin{table}
\begin{center}
\caption[T4]{Total attenuation ($|S(E)|^2$) due to distortion effects in the 
initial and final channels.
}
\vspace{1.0cm}
\begin{tabular}{|c|c|}\hline
Reaction       &      $|S(E)|^2$          \\
\hline
$^{4}$He$(p,K^+)$$^{5}\!\!\!_\Lambda$He$(0s_{1/2})$ & 0.28 \\ 
$^{12}$C$(p,K^+)$$^{13}\!\!\!_\Lambda$C$(0p_{3/2})$ & 0.19 \\
$^{40}$Ca$(p,K^+)$$^{41}\!\!\!_\Lambda$Ca$(0d_{3/2})$ & 0.10 \\
\hline
\end{tabular}
\end{center}
\end{table}

At the same time, one should also note that in the distorted wave treatment
the continuum wave functions are no longer associated  with sharp
momenta but are states with a momentum distribution [see Eqs~(30)-(31)].
This leads to a redistribution of the large momentum transfer 
differently from what is allowed in the plane wave approximation. It could
shift the sensitivity of the model to even lower momenta leading to
enhanced cross sections. The competition between this effect and the 
absorption effect would ultimately decide the role of distortions
in these reactions.  
        
\section{Summary and Conclusions} 

In summary, we have made a study of the $A(p,K^+){_\Lambda}B$ reaction
on  $^{4}$He, $^{12}$C, and $^{40}$Ca targets 
within a fully covariant general two-nucleon mechanism where 
in the initial collision of the incident proton with one
of the target nucleons, $N^*$(1710), $N^*$(1650), and $N^*$(1720) baryonic
resonances are excited which subsequently propagate
and decay into the relevant channel. The initial nucleon-nucleon
collisions are mediated by pion and also by rho and omega exchange
mechanisms. Expressions for the reaction amplitudes are derived in both
distorted wave (DW) and plane wave (PW) approximations. However, the numerical
calculations are performed within the latter approximation which helps 
in understanding the essential features of the $(p,K^+)$ reaction 
without requiring very lengthy and cumbersome computations which are
necessarily involved in the DW method. Wave functions of baryonic
bound states are obtained by solving the Dirac equation with appropriate
scalar and vector potentials.
\begin{figure}
\begin{center}
\mbox{\epsfig{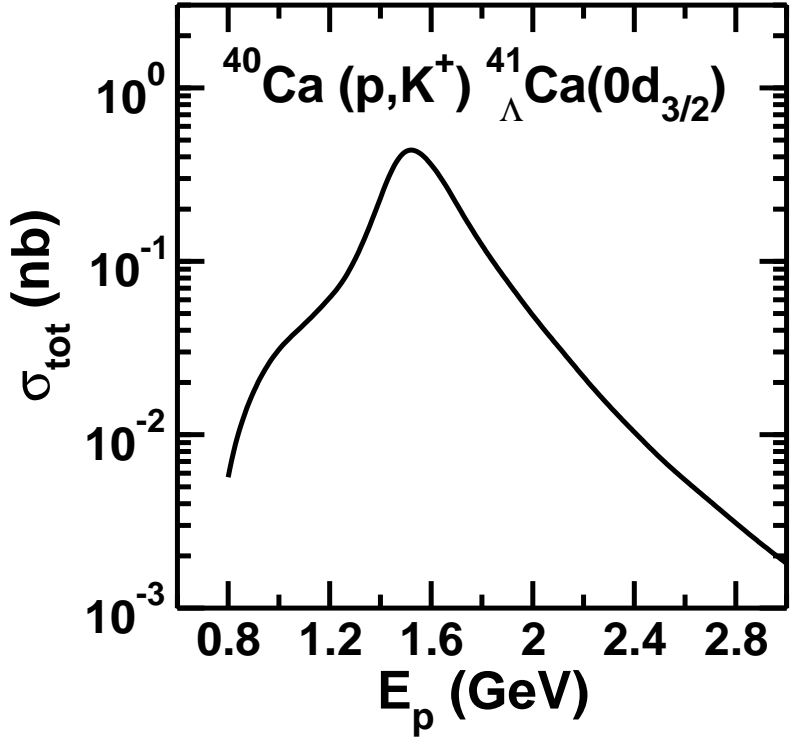}}
\end{center}
\caption{Angle integrated cross section for the 
$^{40}$Ca$(p,K^+)$$^{41}\!\!\!_\Lambda$Ca$(0d_{3/2})$ reaction 
as a function of the incident proton beam energy. 
}
\label{fig:figm)}
\end{figure}

In our model, the $(p,K^+)$ reaction proceeds predominantly via excitation
of the $N^*$(1710) resonant state. For reactions on all the three target
nuclei, the cross sections are dominated by those graphs in which the
intermediate meson originates
from the target nucleon and the projectile nucleon gets excited to the
baryonic resonance state (target emission graph). The one-pion-exchange
processes make up most of the differential cross section at all 
angles. $\rho$ and $\omega$ exchange graphs individually are
relatively more important at backward angles where the momentum transfer
to the nucleus is comparatively larger. The effect of the heavier meson
exchange is to reduce (although very weakly) the pion exchange only cross
sections. However, the  $\rho$ and $\omega$ exchange processes put together
do not appreciably reduce the pion exchange only cross sections even at the 
back angles.

The calculated cross sections are maximum for the hypernuclear state with
the largest orbital angular momentum. This is the typical of the
large momentum transfer reactions~\cite{dov80}. For heavier targets,
the angular distributions for the favored transitions peak at angles
larger than the $0^\circ$ which in contrast to the results of most of
the previous nonrelativistic calculations for this reaction. This reflects
directly the nature of the Dirac spinors for the bound states which involve
several maxima in the region of large momentum transfer. 
In case of the light target $^4$He, however, the differential cross section
still peaks near the zero degree as in this case the momentum transfers are in
the region where the bound state spinors are smoothly decreasing with
momentum. The energy dependence of the calculated total production cross
section follows closely that of the $pp \to p\Lambda K^+$ reaction.

The present results have been obtained using the PW approximation. Effects
of initial and final state interactions have been estimated from an eikonal
approximation for the nuclear elastic scattering interactions of the proton
and kaon. The predicted cross sections are reduced by factors ranging
between 3 - 10.  The reduction is less for the lighter targets. However, 
In the present treatment we considered the absorptive distortions effects
only which influence the absolute magnitudes of the cross sections. In a
more rigorous calculation, distortions may also affect the shapes of
the angular distributions.   

Our results suggest that the study of the $(p,K^+)$ reaction is
attractive as it provides an alternative way to study the 
the spectroscopy of the $\Lambda$ hypernuclear states. This reaction 
should be measurable at the COSY facility in the Forschungszentrum J\"ulich.
The characteristics of these cross sections predicted by us should
be helpful in planning of such experiments. Such experiments may also
lead to more quantitative calculations which should utilize the proper
distorted waves.

\section{ACKNOWLEDGMENTS}

This work has been supported by the Forschungszentrum J\"ulich. We 
would like to thank Pascal M\"uhlich for helping us with the 
calculations of vector meson self-energies.

\appendix
\section{Expression for the amplitudes in distorted wave and plane wave
approximations}

In this appendix we present the final form of the amplitude ($M_{2b}$)
in both full distorted wave as well as plane wave approximations.
\begin{eqnarray}
M_{2b}^{DW}(N^*_{1/2}) & =& -C_{iso}^{2b}\biggl(\frac{g_{NN\pi}}{2m_N}\biggr)
(g_{N_{1/2}^*N\pi}) (g_{N^*_{1/2}\Lambda K^+})
{1 \over {q^2 - m_\pi^2 -\Pi_\pi(q)}} \nonumber \\ & \times &
{1\over {p_{N^*}^2 - (m_{N^*_{1/2}}-i\Gamma_{N^*_{1/2}}/2)^2}} \Biggl[
{\mathscr F}_1 (k_1,k_2) \sum_{m_s} Z_{\ell^\prime j}^{m_j m_s}({\hat p}_1)
Z_{\ell j}^{m_j m_s}({\hat p}_2)^* \nonumber \\ & + &
{\mathscr F}_2 (k_1,k_2)
\sum_{m_s} Z_{\ell j}^{m_j m_s}({\hat p}_1) Z_{\ell^\prime j}^{m_j m_s}
({\hat p}_2)^* \Biggr] \sum_{L_p J_p \ell_K m_K} \sum_{\mu_i} (-)^{\ell_K}
\nonumber \\ & \times &
<L_p 0 1/2 \mu_i | J_p \mu_i > \sqrt{2L_p + 1}\,
{\bf Y}_{\ell_K m_K} (\Omega_K,\Omega_K^\prime)
\nonumber \\ & \times & \Biggl[(c_0 + c_0^\prime E_K)
{\mathscr R}_{\ell_\Lambda L_p \ell_K} Z_{L_p J_P}^{M_p \mu_i}
({\hat p}^\prime_i)
Z_{\ell_\Lambda m_{j_\Lambda} - \mu_i}^{m_{j_\Lambda} \mu_i}
({\hat p}_\Lambda)^* \nonumber \\ & - &
i c_0^\prime {\mathscr R}_{\ell_\Lambda^\prime L_p \ell_K} \{
(p_{K_x}^\prime - i p_{K_y}^\prime)\sqrt{(1/2-\mu_i)(3/2+\mu_i)}
\nonumber \\ & \times & Z_{L_p J_P}^{M_p \mu_i} ({\hat p}^\prime_i)  
Z_{\ell_\Lambda^\prime m_{j_\Lambda} - \mu_i -1 }^{m_{j_\Lambda} \mu_i+1}
({\hat p}_\Lambda)^*  +  (p_{K_x}^\prime + i p_{K_y}^\prime)
\nonumber \\ & \times &
\sqrt{(1/2+\mu_i)(3/2-\mu_i)} Z_{L_p J_P}^{M_p \mu_i} ({\hat p}^\prime_i)
Z_{\ell_\Lambda^\prime m_{j_\Lambda} - \mu_i + 1 }^{m_{j_\Lambda} \mu_i-1}
({\hat p}_\Lambda)^* 
\nonumber \\ & + & 
2p_{K_z}^\prime Z_{L_p J_P}^{M_p \mu_i}({\hat p}^\prime_i)
Z_{\ell_\Lambda^\prime m_{j_\Lambda} - \mu_i}^{m_{j_\Lambda} \mu_i}
({\hat p}_\Lambda)^* \}
\nonumber \\ & - &  c_0^\prime
{\mathscr R}_{\ell_\Lambda L_p^\prime \ell_K} \{
(p_{K_x}^\prime - i p_{K_y}^\prime)\sqrt{(1/2-\mu_i)(3/2+\mu_i)}
\nonumber \\ & \times & Z_{L_p^\prime J_P}^{M_p \mu_i} ({\hat p}^\prime_i)  
Z_{\ell_\Lambda m_{j_\Lambda} - \mu_i -1 }^{m_{j_\Lambda} \mu_i+1}
({\hat p}_\Lambda)^* +  (p_{K_x}^\prime + i p_{K_y}^\prime)
\nonumber \\ & \times & \sqrt{(1/2+\mu_i)(3/2-\mu_i)}
Z_{L_p^\prime J_P}^{M_p \mu_i} ({\hat p}^\prime_i)
Z_{\ell_\Lambda m_{j_\Lambda} - \mu_i + 1 }^{m_{j_\Lambda} \mu_i-1}
({\hat p}_\Lambda)^*
\nonumber \\ & + &
2p_{K_z}^\prime Z_{L_p^\prime J_P}^{M_p \mu_i} ({\hat p}^\prime_i)
Z_{\ell_\Lambda m_{j_\Lambda} - \mu_i}^{m_{j_\Lambda} \mu_i}
({\hat p}_\Lambda)^* \}\nonumber \\ & + &   
i(-c_0 + c_0^\prime E_K)
{\mathscr R}_{\ell_\Lambda^\prime L_p^\prime \ell_K}
 Z_{L_p^\prime J_P}^{M_p \mu_i} ({\hat p}^\prime_i)
Z_{\ell_\Lambda^\prime m_{j_\Lambda} - \mu_i}^{m_{j_\Lambda} \mu_i}
({\hat p}_\Lambda)^* \Biggr],
\end{eqnarray}
where we have defined
\begin{eqnarray}
{\bf Y}_{\ell_K m_K} (\Omega_K,\Omega_K^\prime) & = &
Y_{\ell_K m_K}(\theta_K,0) Y_{\ell_K m_K}^*(\theta_K^\prime,\phi_K^\prime).
\end{eqnarray}
We have, $c_0 = -m_\Lambda + m_{N^*}$, 
and $c_0^\prime = -1$.  Superscript (*) on a function represents the
complex conjugate of the function. In Eq.~(A.1) we have defined
\begin{eqnarray}
Z_{\ell j}^{m_j \mu} ({\hat p}_\lambda) & = & <\ell m_j - \mu 1/2 \mu | j m_j>
 Y_{\ell m_j - \mu}({\hat p}_\lambda),\\
{\mathscr F}_1 (k_1,k_2)& = & 2A g(k_1)f(k_2) + B\xi^\prime(k_1)f(k_2)
                     + B g(k_1) \xi(k_2),\\
{\mathscr F}_2 (k_1,k_2)& = & 2A f(k_1)g(k_2) + B\xi(k_1)g(k_2) + B f(k_1)
                        \xi^\prime (k_2),\\
{\mathscr R}_{\ell_\Lambda L_p \ell_K} & = & f_{\ell_\Lambda} (k_\Lambda)
         F_{L_p J_P}(k^\prime_i,k_i) f_{\ell_K}(k_K^\prime,k_K),\\
{\mathscr R}_{\ell^\prime_\Lambda L_p \ell_K} & = & g_{\ell^\prime_\Lambda}
(k_\Lambda) F_{L_p J_P}(k^\prime_i,k_i) f_{\ell_K}(k_K^\prime,k_K),\\
{\mathscr R}_{\ell_\Lambda L_p^\prime \ell_K} & = & f_{\ell_\Lambda}
(k_\Lambda)G_{L_p^\prime J_P}(k^\prime_i,k_i)
f_{\ell_K}(k_K^\prime,k_K) ,\\
{\mathscr R}_{\ell^\prime_\Lambda L_p^\prime \ell_K} & = &
 g_{\ell^\prime_\Lambda} (k_\Lambda)
       G_{L_p^\prime J_P}(k^\prime_i,k_i) f_{\ell_K}(k_K^\prime,k_K).
\end{eqnarray}
In Eqs.~(A.4) and (A.5), $A = m_N$ and $B = 1$.
There is one more similar term in Eq.~(A.1) involving wave function
$\xi_\ell(k_\Lambda)$ and $\xi_{\ell^\prime}^\prime (k_\Lambda)$
in place of $f_\ell(k_\Lambda)$ and $g_{\ell^\prime}(k_\Lambda)$.
In this term, coefficients $c_0$ and $c_o^\prime$ are replaced by
$c_1$ and $c_1^\prime$ which are -1 and 0, respectively.  

The form of the amplitude $M_{2b}$ in the plane wave approximation is
\begin{eqnarray}
M_{2b}^{PW}(N^*_{1/2}) & =& -C_{iso}^{2b}\biggl(\frac{g_{NN\pi}}{2m_N}\biggr)
(g_{N_{1/2}^*N\pi}) (g_{N^*_{1/2}\Lambda K^+})
{1 \over {q^2 - m_\pi^2 -\Pi_\pi(q)}} \nonumber \\ & \times &
{1\over {p_{N^*}^2 - (m_{N^*_{1/2}}-i\Gamma_{N^*_{1/2}}/2)^2}} \Biggl[
{\mathscr F}_1(k_1,k_2) \sum_{m_s} Z_{\ell^\prime j}^{m_j m_s}({\hat p}_1)
Z_{\ell j}^{m_j m_s}({\hat p}_2)^* \nonumber \\ & + & {\mathscr F}_2(k_1,k_2)
\sum_{m_s} Z_{\ell j}^{m_j m_s}({\hat p}_1) Z_{\ell^\prime j}^{m_j m_s}
({\hat p}_2)^* \Biggr] \Biggl[
Z_{\ell_\Lambda j_\Lambda}^{m_{j_\Lambda} \mu_i} ({\hat p}_\Lambda)^*\,\,T_1
\nonumber \\ & + &
  Z_{\ell_\Lambda^\prime j_\Lambda}^{m_{j_\Lambda} \mu_i}
({\hat p}_\Lambda)^*\,\mu_i \,\,T_4
\nonumber \\ & + &
Z_{\ell_\Lambda j_\Lambda}^{m_{j_\Lambda} \mu_i+1} ({\hat p}_\Lambda)^*
\sqrt{(1/2-\mu_i)(3/2+\mu_i)}\,\, T_2 \nonumber \\ & + & 
Z_{\ell_\Lambda j_\Lambda}^{m_{j_\Lambda} \mu_i-1} ({\hat p}_\Lambda)^*
\sqrt{(1/2+\mu_i)(3/2-\mu_i)}\,\, T_3  
\nonumber \\ & + &
Z_{\ell_\Lambda^\prime j_\Lambda}^{m_{j_\Lambda} \mu_i+1} ({\hat p}_\Lambda)^*
\sqrt{(1/2-\mu_i)(3/2+\mu_i)}\,\, T_5 \nonumber \\ & + & 
Z_{\ell_\Lambda^\prime j_\Lambda}^{m_{j_\Lambda} \mu_i-1} ({\hat p}_\Lambda)^*
\sqrt{(1/2+\mu_i)(3/2-\mu_i)}\,\, T_6 \Biggr],
\end{eqnarray}  
where we have defined
\begin{eqnarray}
T_1 & = & (c_0 + c_0^\prime\,E_K)\,f(k_\Lambda) - c_0^\prime {\frac{\hbar c}
{E_i + m_N}}\, p_i p_K\, sin\,\theta_K\,f(k_\Lambda) \nonumber \\ & + &  
(c_1 + c_1^\prime\,E_K)\,\xi(k_\Lambda)c_1^\prime {\frac{\hbar c}
{E_i + m_N}}\, p_i p_K\, cos\,\theta_K \,\xi(k_\Lambda),\\
T_2 & = & c_0^\prime {\frac{\hbar c}{E_i + m_N}} \,
p_i p_K\, sin\,\theta_K\,f(k_\Lambda) \nonumber \\ & + &
 c_1^\prime {\frac{\hbar c}{E_i + m_N}}
\,p_i p_K\, sin\,\theta_K \,\xi(k_\Lambda),\\
T_3 & = & -T_2, \\
T_4 & = & -2i\, c_0^\prime\,p_K\,cos\,\theta_K\, g(k_\Lambda) +
 2i\,{\frac{\hbar c}{E_i + m_N}}\,
(-c_0 + c_0^\prime\,E_K)\,p_i\,g(k_\Lambda)\\
& - & 2i\, c_1^\prime\,p_K\,cos\,\theta_K\, \xi^\prime(k_\Lambda) +
2i\,{\frac{\hbar c}{E_i + m_N}}\,(-c_1 + c_1^\prime\,E_K)\,p_i\,\xi^\prime
(k_\Lambda),\\
T_5 & = & -i\, c_0^\prime\,p_K\,sin\,\theta_K\, g(k_\Lambda) -
          -i\, c_1^\prime\,p_K\,sin\,\theta_K\, \xi^\prime(k_\Lambda),\\
T_5 & = & T_6
\end{eqnarray}
The coefficients $c_0$, $c_0^\prime$, $c_1$ and $c_1^\prime$  are the same
as defined above.

The amplitude for the graph 2(c) can be written in the analogous way. 
Those involving the excitation of other two resonances and vector 
meson exchanges have similar structure even though they may be a bit
more complicated.
\section{Continuum distorted wave functions in momentum space}

In this appendix we present a short derivation of Eqs.~(30)-(31).

The incident and outgoing particles are described by stationary continuum 
wave functions $\Psi(r_\alpha) = \psi_\alpha({\bf p}_\alpha,{\bf r}_\alpha)
\,exp(-iE_\alpha t)$ where $\alpha$ = $i$ represents the incident channel 
and $K$ the outgoing one. $\psi_\alpha({\bf p}_\alpha,{\bf r}_\alpha)$
are given, in the partial wave representation, by 
\begin{eqnarray}
\psi_i^{(+)}({\bf p}_i,{\bf r}_i) & = & \sum_{J_p L_p M_p}
i^{L_{p}}<L_{p} M_{p}-\mu 1/2 \mu|J_{p} M_{p}> 
      Y^*_{L_{p}M_{p}-\mu}(\hat{p}_i) \nonumber \\ & \times &
{{F^C_{L_{p} J_{p}}(k_i,a_i){\mathscr Y}_{L_{p} 1/2 J_{p}}^{M_{p}}
                 ({\hat r}_i)} \choose
{G^C_{L^\prime_{p} J_{p}}(k_i,a_i)
{\mathscr Y}_{L^\prime_{p} 1/2 J_{p}}^{M_{p}}
                     ({\hat r}_i)}}, \\
\phi_{K}^{(-)*}({\bf p}_K,{\bf r}_K) & = & \sum_{\ell_K m_K} i^{-\ell_K}
 Y_{\ell_K m_K} ({\hat p}_K) f_{\ell_K}(k_K,a_K)
 Y_{\ell_K m_K}^*({\hat r}_K),
\end{eqnarray}
where various symbols have the same meaning as those described in the
main text. Having chosen a momentum space formalism, we need to calculate
the Fourier transforms
\begin{eqnarray}
\Psi_\alpha (p^\prime,p_\alpha) & = & \int\frac{d^4r_\alpha}
{(2\pi)^4} e^{ip^\prime r_\alpha} \Psi_\alpha(r_\alpha) ,
\end{eqnarray}
depending on the unconstrained Fourier four momentum $p^\prime =
(p^\prime_0,{\bf p}^\prime)$ and the asymptotic on-shell momenta 
$p_\alpha = (E_\alpha,{\bf p}_\alpha)$. The time integral in Eq. (B.3),
can be trivially performed, leading to 
\begin{eqnarray}
\Psi_\alpha (p^\prime,p_\alpha) & = & \delta(p^\prime_0 -E_\alpha)
\int\frac{d^3{\bf r}_\alpha}{(2\pi)^3} e^{-i{\bf p}^\prime \cdot {\bf r_\alpha}}
\psi_\alpha({\bf p}_\alpha,{\bf r}_\alpha).
\end{eqnarray}
Further integrations in Eq.~(B.4) can be carried out by using the
relation
\begin{eqnarray}
e^{-i{\bf p^\prime}\cdot{\bf r}_\alpha} & = & 4\pi \sum_{L M} i^{-L}
 j_L(k^\prime a_\alpha)\,\, Y_{L M}({\hat {p}}^\prime)\,\,
 Y_{L M}^*({\hat r}_\alpha), 
\end{eqnarray}
which lead to  Eqs.~(30)-(31).

\end{document}